\documentclass[]{spie}  %>>> use for US letter paper
\usepackage{amsmath,amsfonts,amssymb}
\usepackage{graphicx}
\usepackage{threeparttable}
\usepackage[colorlinks=true, allcolors=blue]{hyperref}

\title{Generic integration-time hysteretic sensor response revealed in CCD photon transfer statistics and implications for high fidelity point-spread function retrieval: \\
connecting recorded signal to incident flux distribution in high dynamic range exposures}

\title{Extracting incident flux from recorded signal in high dynamic range exposures: high fidelity point-spread function retrieval in the presence of integration time hysteretic sensor response}

\title{High fidelity point-spread function retrieval in the presence of electrostatic, hysteretic pixel response}

\author[a]{Andrew Rasmussen}
\author[b]{Augustin Guyonnet}
\author[c]{Craig Lage}
\author[b]{Pierre Antilogus}
\author[b]{Pierre Astier}
\author[d]{Peter Doherty}
\author[a]{Kirk Gilmore}
\author[e]{Ivan Kotov}
\author[f]{Robert Lupton}
\author[e]{Andrei Nomerotski}
\author[e]{Paul O'Connor}
\author[d]{Christopher Stubbs}
\author[c]{Anthony Tyson}
\author[g]{Christopher Walter}
\affil[a]{SLAC National Accelerator Laboratory, Menlo Park, CA, United States}
\affil[b]{LPHNE/IN2P3/CNRS, UPMC, France}
\affil[c]{University of California at Davis, Davis, CA, United States}
\affil[d]{Harvard University, Cambridge, MA, United States}
\affil[e]{Brookhaven National Laboratory, Upton, NY, United States}
\affil[f]{Princeton University, Princeton, NJ, United States}
\affil[g]{Duke University, Durham, NC, United States}

\authorinfo{Send correspondence to A.R. -- E-mail: arasmus@slac.stanford.edu; Tel: +1 650 926 2794; Fax: +1 650 926 5566\\
Full electronic version of this paper with color figures is available on arXiv.org [astro-ph.IM]}

\begin{document} 
\maketitle

\begin{abstract}
We employ electrostatic conversion drift calculations to match CCD pixel signal covariances observed in flat field exposures acquired using candidate sensor devices for the LSST Camera~\cite{Kurita_2016,Kahn_2016}. We thus constrain pixel geometry distortions present at the end of integration, based on signal images recorded. We use available data from several operational voltage parameter settings to validate our understanding. Our primary goal is to optimize flux point-spread function (FPSF) estimation quantitatively, and thereby minimize sensor-induced errors which may limit performance in precision astronomy applications.  We consider alternative compensation scenarios that will take maximum advantage of our understanding of this underlying mechanism in data processing pipelines currently under development.

To quantitatively capture the pixel response in high-contrast/high dynamic range operational extrema, we propose herein some straightforward laboratory tests that involve altering the time order of source illumination on sensors, within individual test exposures. Hence the word {\it hysteretic} in the title of this paper.
\end{abstract}


%\section{INTRODUCTION}
%\label{sec:intro}  % \label{} allows reference to this section

%Begin the Introduction below the Keywords. The manuscript should not have headers, footers, or page numbers. It should be in a one-column format. References are often noted in the text and cited at the end of the paper.

%Here is a test for citations: Rasmussen~\cite{Rasmussen_paccd_2015,Rasmussen_paccd_2014,Rasmussen_spie_2014}. Kurita~\cite{Kurita_2016,Kahn_2016}. Lage~\cite{Lage_lastkpc_2015} and Bradshaw~\cite{Bradshaw_paccd_2015}. Guyonnet~\cite{Guyonnet_2015}! Downing~\cite{Downing_2006}.. Gruen~\cite{Gruen_jinst_2015}.

\section{INTRODUCTION}
\label{sec:intro}  % \label{} allows reference to this section

The literature already includes several instances of how flat field correlations may be used to correct or compensate astronomical data for the so called {\it brighter-fatter} (BF) effect\cite{Antilogus_paccd_2014,Guyonnet_2015,Gruen_jinst_2015}.

In a separate work, Niemi {\it et al.}\cite{Niemi_2015} opted to favor the information available from {\it direct}, focused spot measurements over the {\it indirect} information from flat field correlations, and that the flux level dependence to measured spot sizes was described as an {\it intrinsic CCD PSF} that depends on intensity and wavelength only. A key piece of the puzzle that conflicts with this picture is that the BF effect nearly vanishes when a single pixel's center is illuminated with a spot of sub-pixel diameter\cite{Miyazaki:privcomm}: the instrument's signature that contributes to systematic errors in turn must depend on the incident flux distribution at the sensor's entrance window as well as the instantaneous recorded signal distribution as illumination progresses. Consequently, we argue 
that contributions from the instrument cannot be separated from the contributions of the incident flux,
%there is no credible way to separate the instrument's contributions from the incident flux, 
unless the integration of the recorded image is also considered in the process.

In this contribution, we approach the issue by calculating changes to pixel areas based on families of electrostatic solutions to Poisson's equation in the drift region\cite{Rasmussen_paccd_2015}. A separate effort, not discussed in detail here, applies a Poisson solver to sensor's semiconductor properties informed by detailed fabrication steps and lithographic information, provided by the vendor\cite{Lage_lastkpc_2015}. 

\section{Dynamic changes to pixel areas and their relationship to measured covariances}\label{sec:areas_covariances}

Direct pixel boundary calculations, if they indeed reproduce available characterization data, are likely to be preferable to pixel border shift models because they provide a two-dimensional point-to-pixel partition (they do not over- or under-count area elements), and also offer chromaticity information (pixel boundaries as a function of conversion depth) in the dynamic pixel geometric response.  Direct boundary calculations properly handle built-in nonlinearities in pixel geometries as aggressor signals, due to accumulated conversions, approach full well. In the context of the rolled-up model we describe, the recorded signal distribution of an image is used to calculate the self-consistent, distorted pixel boundaries in effect at the end of the exposure. Better constraints on the incident flux distribution would naturally result from knowledge of those boundaries.

In the following, we provide some calculations that extend work discussed in the aforementioned references to connect pixel correlations to their theoretical signal level dependence, whether for flat field or focused spot applications.  We compute pixel area response to aggressor signal level.

\subsection{Flat field statistical fluctuations and signal expectation values as a function of lag}\label{ssec:ff_statistical}

In this discussion, the aggressor is a statistical fluctuation $\zeta$ in a recorded flat field image, that occurs about the mean $\mu$ and induces changes in neighboring pixels' areas. A neighboring pixel is indicated by its lag from the aggressor using two indices, $ij$, where $i$ and $j$ are the lags along the serial and parallel directions, respectively. Because we calculate area variations due to drift during collection and not transfer statistics due to trap populations and channel occupancy, full descriptions of area variations are captured by considering only positive $i$ and $j$. Correspondingly, $ij=00$ indicates the aggressor pixel and $q_{00}$ is the charge signal accumulated there. 

Considering the {\it direct aggressor--victim channel} only, a nearby pixel with lag $ij$ has an area at the end of integration: 
\begin{equation}
\Delta \ln a_{ij}\left(q_{00}=\mu+\zeta | \mu \right)\approx { d \ln a_{ij} \over dq_{00}} \zeta,
\label{eq:approx}
\end{equation}
and on average this pixel would contain a signal level that is systematically biased by the area distortion. While the area distortion is zero at the beginning of the integration ($\Delta a_{ij}=0$~@~$t=0$), it should be finite by integration end. Averaged over all possible trajectories $\Delta a_{ij}(t)$, the influence of the statistical fluctuation in pixel $ij=00$ on the bias $\left< \Delta q_{ij}\right>$  is readily isolated:  
\begin{equation}
\left<\Delta q_{ij}\left(q_{00}|\mu\right)\right> = 
{\mu \over 2}\left( \exp\left( \Delta \ln a_{ij}\left( q_{00}|\mu \right) \right) - 1 \right)
\approx
{1 \over 2} \mu \zeta {d \ln a_{ij} \over dq_{00}}=
\mu\zeta{d\ln \bar{a}_{ij}\over dq_{00}}
\label{eq:dq_ij}
\end{equation}
where for convenience, we also use the exposure time averaged pixel area $\bar{a}_{ij}$.
The expression for the covariance $\mathrm{Cov}_{ij}$ may also be simplified by using the same approximation, and the variance $\mathrm{Var}\equiv \mathrm{Cov}_{00}$ appears:\footnote{For consistency in nomenclature, we define the zero lag covariance to be equal to the variance: 
\begin{equation*}
\mathrm{Cov}_{00} = {\sum_{kl}\zeta_{kl}\Delta q_{k+0,l+0} \over\sum_{kl}1} = {\sum_{kl}\zeta_{kl}^2 \over\sum_{kl}1}\equiv \mathrm{Var}.
\end{equation*}
}

\begin{align}
\mathrm{Cov}_{ij} &= 
{\sum_{kl} \zeta_{kl} \left< \Delta q_{k+i,l+j} \right> \over \sum_{kl} 1}
=
{\mu \over 2} {\sum_{kl} \zeta_{kl} \left( \exp\left( \Delta \ln a_{k+i,l+j} \left( q_{kl} | \mu \right) \right) - 1 \right) \over \sum_{kl}1}\label{eq:covij_approxa}
\\
&\approx 
{\mu \over 2} {\sum_{kl} \zeta_{kl}^2 {d\ln a_{k+i,l+j}\over dq_{kl}} \over \sum_{kl}1}
=
{\mu\over 2} \mathrm{Cov}_{00} {d\ln a_{ij}\over dq_{00}}
.
\label{eq:covij_approxb}
\end{align}
The correlations are then the covariances divided by the variance, and following through with the same approximation, they are proportional to the area response $d\ln a_{ij}/dq_{00}$ and the scaling term $\mu/2$:
\begin{align*}
\mathrm{Corr}_{ij} \equiv {\mathrm{Cov}_{ij} \over \mathrm{Cov}_{00}} \approx {\mu \over 2} {d\ln a_{ij} \over dq_{00}}
\end{align*} 

Now irrespective of finite correlations in the flat fields, Poisson statistics are recovered by re-binning images (in the case of data frames), or in the case of finite pixel area distortions:
\begin{align*}
{\sum_{kl}\left(\zeta_{kl}\sum_{ij}\zeta_{k+i,l+j}\right)\over \sum_{kl}\sum_{ij}1}
=
{\sum_{kl} \zeta_{kl}^2\over \sum_{kl}1}+{\sum_{kl}\sum_{ij \neq 00}\zeta_{kl}\zeta_{k+i,l+j} \over \sum_{kl}\sum_{ij\neq 00}1}=\mathrm{Cov}_{00}+\sum_{ij\neq 00}\mathrm{Cov}_{ij}\equiv \mu.
\end{align*}
Because the expressions for $\mathrm{Cov}_{ij}$ are symmetric under exchange $i\rightarrow -i$ and $j\rightarrow -j$, the above expression may be further simplified to include only the unique quantities $\mathrm{Cov}_{ij}$ for $i \ge 0,j \ge 0$:
\begin{equation}
\mu=
\sum_{ij}\mathrm{Cov}_{ij}=
\mathrm{Cov}_{00} + 2\sum_{i\ge 0,j\ge 0}\left(2-\left[\delta_{0i}+\delta_{0j}\right]\right)
\mathrm{Cov}_{ij},
\label{eq:poisson_conservation}
\end{equation}
where $\delta_{ij}$ is the Kronecker delta. Finally, area is conserved, such that any area lost (or gained) by pixel $ij=00$ is recovered (or ceded) by others:
\begin{equation}
\sum_{ij}\Delta a_{ij} = 0 = 
\left(\exp\left(\Delta \ln a_{00}\left(q_{00}|\mu\right)\right)-1\right)+
\sum_{ij\neq 00}\left(\exp\left(\Delta \ln a_{ij}\left(q_{00}|\mu\right)\right)-1\right).
\label{eq:area_conservation}
\end{equation}

Equations \ref{eq:poisson_conservation} \& \ref{eq:area_conservation} provide a way to express the shape of the mean-variance curve. Solving for $\mathrm{Cov}_{00}$:
\begin{align}
\mathrm{Cov}_{00} &= 
%\mu\left(1-\sum_{ij\neq 00}\mathrm{Cov}_{ij}\right)=
\mu - \sum_{ij\neq 00}\mathrm{Cov}_{ij} =
{\mu \over \left(1+\sum_{ij\neq 00}{\mathrm{Cov}_{ij}\over \mathrm{Cov}_{00}}\right)}
\label{eq:cov00a}
\\
&\approx
{\mu\over 1+{\mu\over 2}\sum_{ij\neq 00}{d\ln a_{ij}\over dq_{00}}} = 
{\mu\over 1-{\mu\over 2}{d\ln a_{00}\over dq_{00}}} = \mu\sum_{n=0}^\infty\left({\mu\over 2} {d\ln a_{00}\over dq_{00}}\right)^n.
\label{eq:cov00b}
\end{align}

Evidently, the observed deviation from Poissonian behavior in the mean-variance curve indicates ${d\ln a_{00} \over dq_{00}}<0$ \cite{Antilogus_paccd_2014,Rasmussen_paccd_2015} and the approximate relation is valid if $|{\mu\over 2}{d \ln a_{00}\over q_{00}}| \ll 1$. In general however, Equation~\ref{eq:cov00a} should remain valid independent of this assumption. The general form of this is Equation~\ref{eq:cov00}. Routine, accurate gain determination may be enabled by fitting functions of this form to gain-variance measurement pairs that show this curvature~\cite{OConnor_2016}.

\subsection{Application to measured covariances}\label{ssec:application_to_measured_covariances}
We apply the equations above to the specific regime where statistical fluctuations $\zeta$ are much smaller than (and tied to) the flat field flux $\mu$. Recall that in the Poisson limit, $\zeta^2 \equiv \mu$, but this is apparently not correct in actual photon transfer curves. High-quality flat field data sets can be used to generate a pattern of lag ($ij$) specific covariances $\mathrm{Cov}_{ij}$, which are in turn converted into correlations via $\mathrm{Corr}_{ij}\equiv \mathrm{Cov}_{ij}/\mathrm{Cov}_{00}$.  Generally, $\mathrm{Cov}_{ij\ne 00}$ scale as $\mu^2$, while $\mathrm{Corr}_{ij\ne 00}$ scale as $\mu$. From the modeling side, a statistical fluctuation translates to an aggressor amplitude $\bar{p}$ which in turn produces the pattern of area distortions $\Delta a_{ij}(\bar{p}|\mu)$ that govern the biases in the expression for $\mathrm{Cov}_{ij}$ (Eq.~\ref{eq:dq_ij}). Taking the {\it rms} exposure averaged aggressor to be $\bar{p}\equiv z_{chan}\zeta \mathrm{q}_e$, we write the following:

\begin{eqnarray}
\mathrm{Cov}_{ij} &=& \mu\zeta \Delta a_{ij}(\bar{p})\nonumber\\
\mathrm{Cov}_{00} &=& \zeta^2 = {\mu \over 1- {\mu\over\zeta}\Delta a_{00}(\bar{p})}\label{eq:cov00}\\
\mathrm{Corr}_{ij} &=& {\mu \over \zeta} \Delta a_{ij}(\bar{p}) = \left(\zeta - \mu \Delta a_{00}(\bar{p})\right)\, \Delta a_{ij}(\bar{p})\label{eq:corij}.
\end{eqnarray}
In the above, $\mathrm{Corr}_{ij}$, $\zeta$ and $\mu$ are measurements and $\Delta a_{ij}(\bar{p})$ are compiled from results of the drift calculation.

\begin{figure} [ht]
\begin{center}
\begin{tabular}{cc} %% tabular useful for creating an array of images 
\includegraphics[height=8cm,width=8cm]{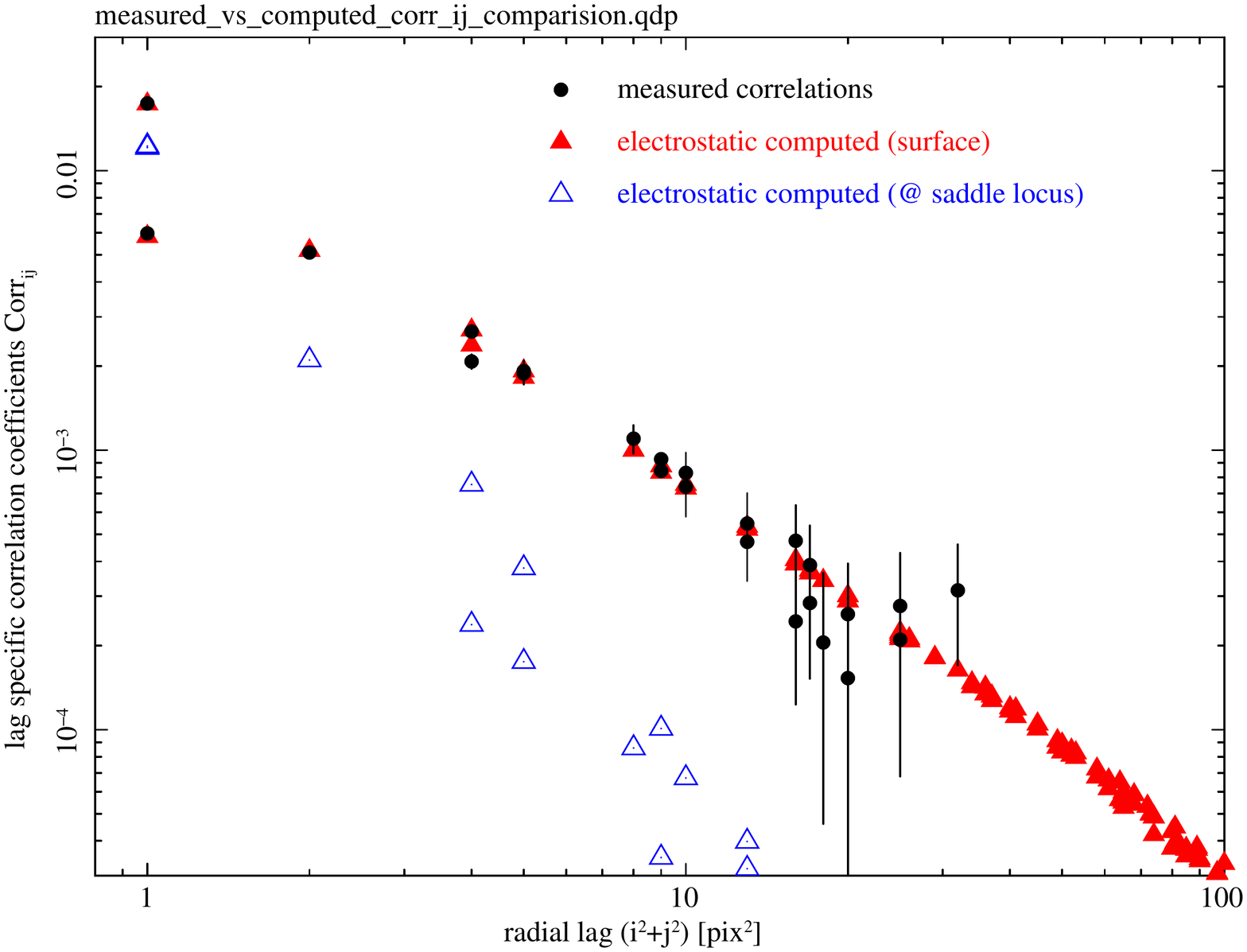}&
\includegraphics[height=8cm,width=8cm]{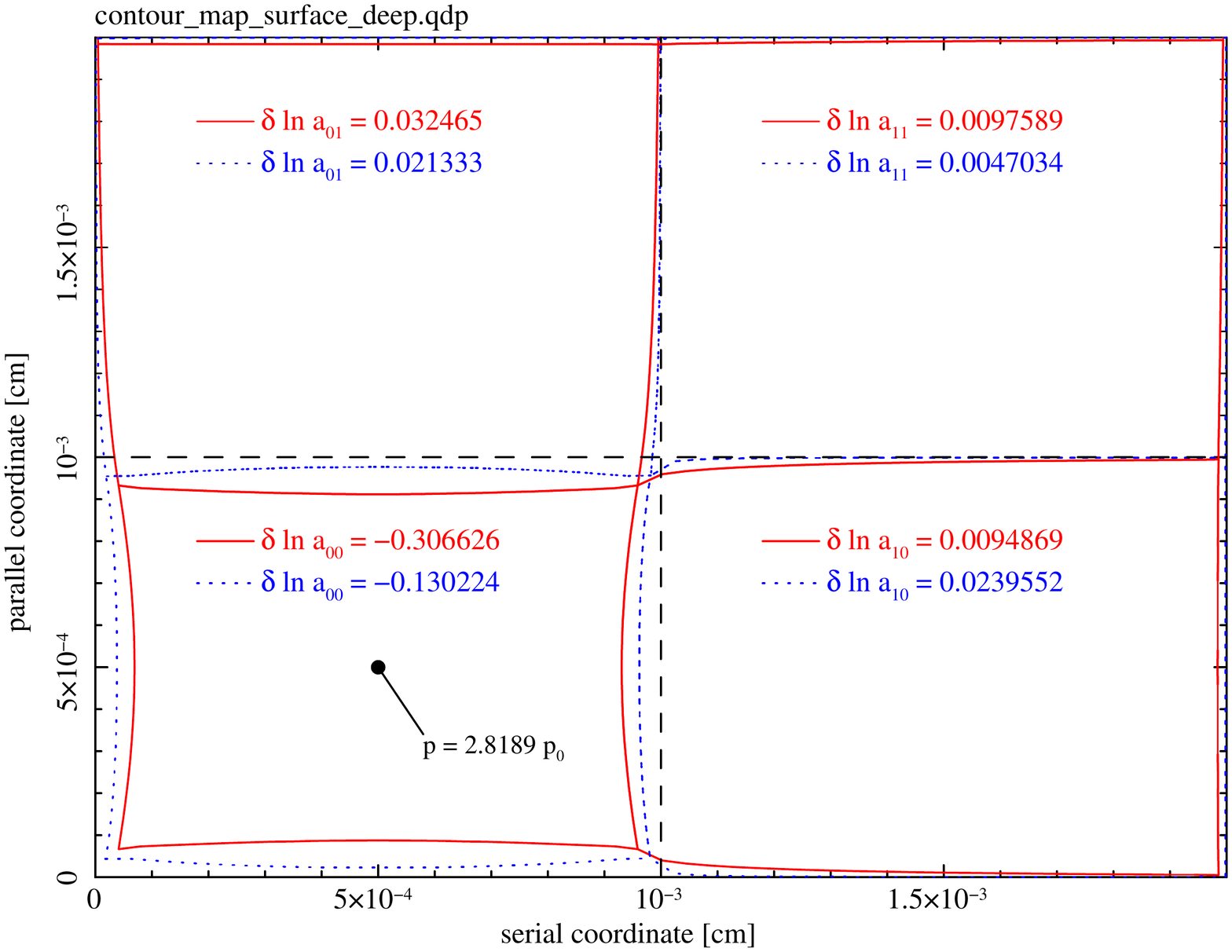}
\end{tabular}
\end{center}
\caption{ \label{fig:pixel_area_model} 
Pixel area variations. Left: a comparison of the best-fit electrostatic drift model to measured correlation coefficients for a specific flat field illumination. Measurements of $\mathrm{Corr}_{ij}$ are the dots (black) with error bars, filled triangles (red) are for shallow conversions, 
and open triangles (blue) are limiting cases for the conversions occurring near the saddle locus along the drift lines. Any observed chromaticity in the flat field correlations should produce numbers that lie somewhere between the shallow and deep limits corresponding to the same lag $ij$.
Best-fit parameters, along with quantities that affect the interpretation of this model are given in Table~\ref{tab:bestfit_pars}. Right: a graphical representation of the same electrostatic model's pixel boundary distortions, for the nearest neighbors (the aggressor is denoted lag $ij=00$). The solid (red) lines show the pixel boundaries for cold electrons very close to the backside surface, while the dotted (blue) lines show a two-dimensional projection of the {\it saddle locus}, where adjacent drift lines diverge to feed the channels belonging to adjacent pixels. The dashed (black) lines, together with the plot frame, show positions of the undistorted pixel boundaries, i.e., for zero aggressor amplitude. For the purpose of this graphic, the aggressor dipole moment amplitude $p$ was increased by a factor of $\sim 500$ as compared to the best-fit on the left (while holding all other parameters fixed).
}
\end{figure}

A measured correlation pattern $\mathrm{Corr}_{ij}$, together with estimates for $\mu$ and $\zeta$ for a specific flat field illumination were used to fit an electrostatic drift model for its undetermined parameters. The data were acquired from a candidate sensor prototype for the LSST Camera manufactured by e2v: it was described previously in Ref.~\citenum{Guyonnet_2015}. $\chi^2$ was minimized using the Nelder-Mead method with results shown in Figure~\ref{fig:pixel_area_model} (left). The best-fit parameter list is given in Table~\ref{tab:bestfit_pars}.  Four free parameters were jointly estimated in the process.  The area distortion model appears to provide enough detail to reproduce the measured, anisotropic correlation pattern and falloff with separation. In addition to constraining magnitudes of the periodic barrier dipole moments, the fit also favors a specific value for the impurity concentration in the silicon bulk, $N_a$. The goodness of fit was acceptable when using estimated $\mu$ and $\zeta$, so the process completed without invoking a gain error parameter in Eq.~\ref{eq:corij}. 

A secondary result of this fitting procedure is a value for the channel depth. For $\mu\sim65\,\mathrm{ke^-}$, we find $z_{chan}=\bar{p}/(\zeta\mathrm{q_e})\approx 2.37\mu\mathrm{m}$. This result is shown below (\S\ref{ssec:z_twiddle}) to constrain other physical properties of the sensor.

\begin{table}[ht]
\caption{
\label{tab:bestfit_pars}
Parameter list for the best-fit electrostatic drift model (for cold carriers)}
\begin{center}
\begin{threeparttable}
\begin{tabular}{|c||c|c|c|}
\hline
parameter & value & units & comments\\
\hline\hline
$N_a$ & $1.11\times 10^{12}$ & $\mathrm{cm}^{-3}$&acceptor density in depleted Si\\
$t_{Si}$ & $100$ & $\mu$m & sensor thickness (fixed) \\
$BSS$ & $-78$ & Volts & backside bias (fixed)\tnote{a}\\
$\xi_{cs}$ & $12.407$ & $\xi_0$\tnote{b} & channel stop 2-D dipole moment \\
$\xi_{ck}$ & $2.6425$ & $\xi_0$\tnote{b} & clock barrier 2-D dipole moment \\
$\bar{p}$ & $0.0057208$ & $p_0\tnote{c}$ & aggressor dipole moment\tnote{d}\\
\hline
$\mu$ & 65230 & $\mathrm{e}^-$ & mean signal level in flat \\
$\zeta^2$ & 58429 & $(\mathrm{e}^-)^2$ & variance in flat\tnote{d} \\
\hline
\end{tabular}
\begin{tablenotes}
\item[a] constrained by measured X-ray diffusion variation with BSS on a similar device
\item[b] $\xi_0\equiv 10^{-6}\,\mathrm{q}_e$
\item[c] $p_0\equiv 10^{5}\,\mu\mathrm{m}\,\mathrm{q}_e$
\item[d] exposure averaged, {\it rms} aggressor moment is $\bar p \equiv z_{chan}\,\zeta\,\mathrm{q}_e$
\end{tablenotes}
\end{threeparttable}
\end{center}
\end{table}

\section{Tunable electrostatic influences to pixel areas}\label{sec:electrostatic_influences}
We have a starting point for more detailed modeling. It is a relatively straightforward task to reproduce fixed pattern features seen in the sensors that would be categorized as cosmetics -- but can in fact be traced to pixel area distortions. In previous work we have demonstrated success in reproducing observed features seen in flat field illumination via electrostatic modeling: edge rolloff, midline charge redistribution, bright and dark column pairs identified as {\it tearing} features~\cite{Rasmussen_paccd_2014}, and {\it bamboo}~\cite{Rasmussen_spie_2014}, each with self-consistent pixel shifts and elongations that accompany the pixel area distortions revealed in the flat fields~\cite{Rasmussen_paccd_2015}. The important difference between this and prior work is that the statistical properties of the flat field illumination, $\mathrm{Cov}_{ij}$, are also reproduced at the same time.

\begin{figure} [ht]
\begin{center}
%\begin{tabular}{cc} %% tabular useful for creating an array of images 
\begin{minipage}[h]{7.5cm}
\vspace{0.4cm}
\centering
\includegraphics[height=7.5cm,width=7.5cm]{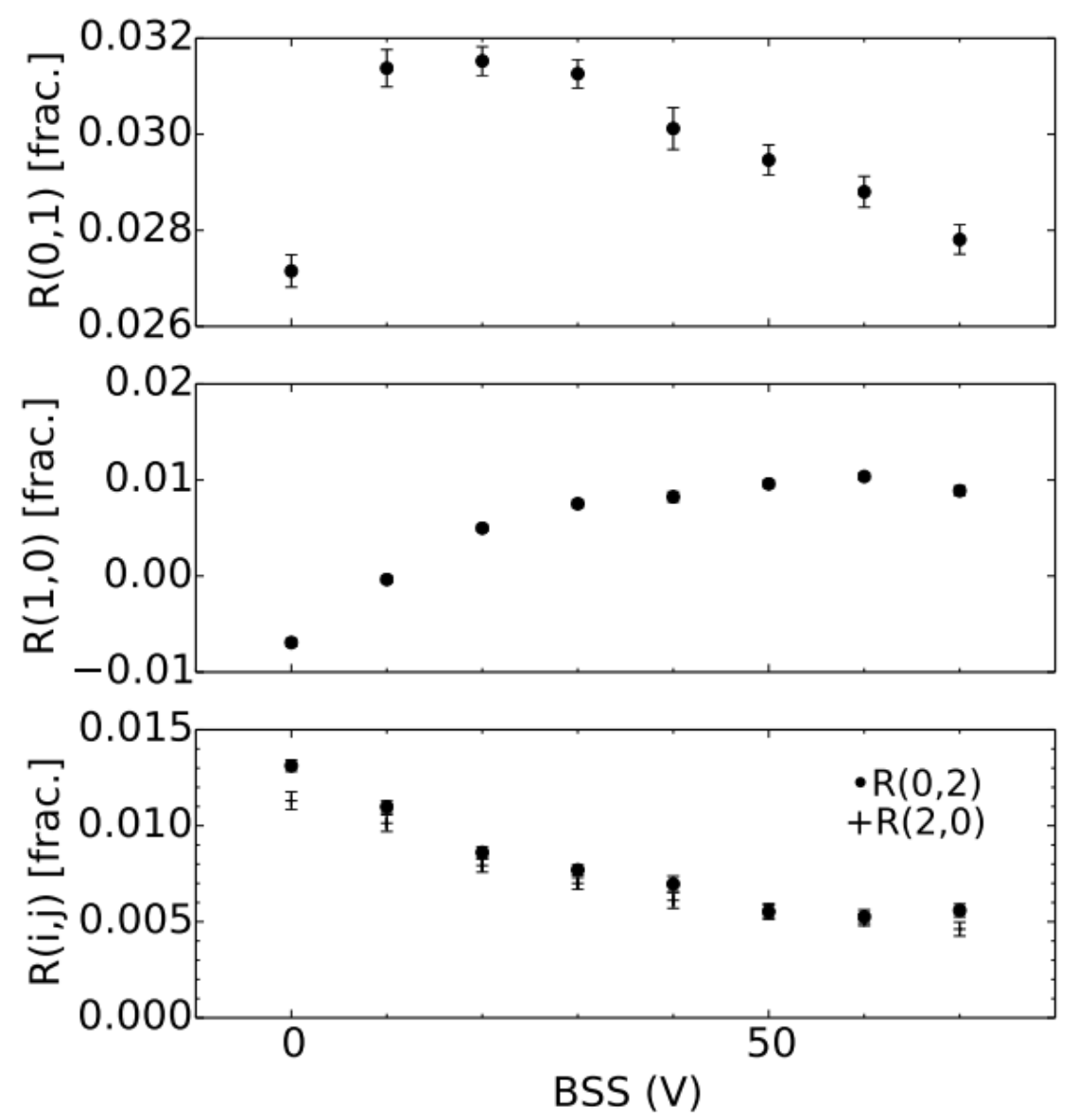}
\end{minipage}
\begin{minipage}[h]{8.5cm}
\centering
\includegraphics[height=8.8cm,width=8.0cm]{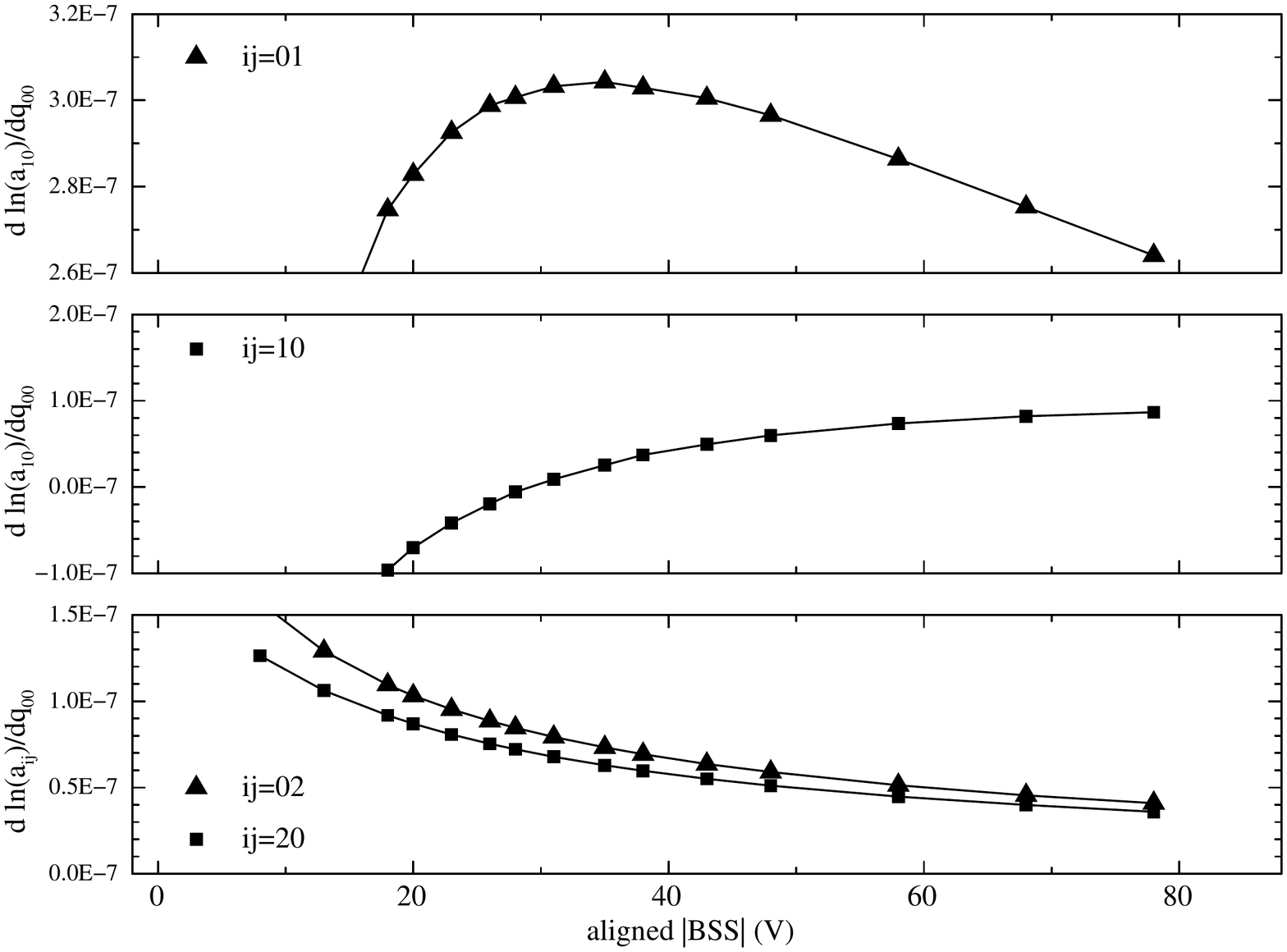}
\end{minipage}
%\end{tabular}
\end{center}
\caption{
\label{fig:bss_dep} 
A blind, qualitative comparison between measured correlation coefficients (left) and computed area distortions (right) as the backside bias is varied. Left: this plot was reproduced from Fig.~8 of Guyonnet {\it et al.}~\cite{Guyonnet_2015} and represents flat field correlations measured for several lags $ij \in \lbrace 01,10,02,20\rbrace$ for $\mu\sim 100\,\mathrm{ke^-}$. Right: computed per-electron area distortions for the same lag selection, for the model summarized in Table~\ref{tab:bestfit_pars}, fit for $|BSS|=70$V (``aligned'' $|BSS|=-78$V\cite{Rasmussen_spie_2014}). While even the $\mathrm{Corr}_{ij}$ pattern shown here (for $|BSS|=70$V) isn't entirely consistent with the $\mathrm{Corr}_{ij}$ pattern shown in Figure~\ref{fig:pixel_area_model} (left) collected for $\mu\sim65\,\mathrm{ke^-}$, this imperfect comparison shows that counterintuitive results seen in the data are readily reproduced in the drift model. These include changes in sign of $\mathrm{Corr}_{ij}$ and their derivatives with respect to $|BSS|$. The ordinate scales on the right hand plots were adjusted to compare directly to the corresponding plots on the left.
}
\end{figure} 

The best-fit electrostatic parameters given in Table~\ref{tab:bestfit_pars} can represent fiducial performance, and small changes in their values will consequently affect the dynamic, hysteretic response of the sensor. For example, a change in clock rail voltage differential would induce a proportional change in $\xi_{ck}$, a change in backside bias ($BSS$) would induce a change in $\vec{E}^{BD}(z)$ according to Appendix~\ref{apxsec:estat_summary}, Equation~\ref{eq:ebd}, and operation with {\it tearing} present (betrayed by darker column pairs straddling segment boundaries in flat field response) would alter $\xi_{cs}$\cite{Rasmussen_paccd_2014} and boost the BF effect due to reduced barriers there. Moreover, parameterization of the {\it aggressor moment} $p$, its dependence on signal $\zeta$ and the geometric response of pixel boundaries (see \S\ref{sec:electrostatic_influences}) can provide an informed process by which recorded images can be used to constrain flux distributions incident at the sensor entrance window. A check for how $\mathrm{Corr}_{ij}$ vary with $BSS$ between measurement and calculation is given in Figure~\ref{fig:bss_dep}. It demonstrates, via a blind test, that counterintuitive dependencies in the measured $\mathrm{Corr}_{ij}$ are reproduced using the simple, far-field approximate, multipole expansion drift model that explicitly satisfies Poisson's equation.  The Dirichlet solution approach\cite{Lage_lastkpc_2015} to describe the sensor's photosensitive volume from the clocks and channel stops toward the backside surface, may still be required to accurately model or understand other details of sensor operation, but these are not defined or addressed here. This serves as a proof of concept that the drift model may already be adopted for more widespread application in data analysis to reduce sensor systematics.

\subsection{The brighter-fatter template}\label{ssec:bf_template}
Several efforts could benefit from this detailed and robust modeling. These include astronomical data reduction pipelines and simulation tools (based on image, table, or ray tracing). Efficient implementations require that results of the time consuming electrostatic drift calculation be ported to faster simulation or analysis frameworks. This was discussed previously (Ref.~\citenum{Rasmussen_paccd_2015}, \S5.3) but is briefly summarized and expanded for completeness in Appendix~\ref{apxsec:template}. A single calculation result, in the format of a {\it BF template} could be applied over broader range of aggressor amplitudes, indeed up to the canonical {\it full well depth} for these sensors, using the linear perturbation model. The work addressed in \S\ref{ssec:da_vs_p} tests this notion.

% This is consistent with the {brighter-fatter} correction implemented in some existing astronomical pixel data pipelines~\cite{Gruen_jinst_2015,Lupton:privcomm} that perform a redistribution of recorded signal using proxies for local fluxes and flat field covariances .

% \subsection{Correlation coefficients vs. sensor backside bias}\label{ssec:corr_vs_BSS}
\subsection{Pixel area dependence on aggressor dipole moment}\label{ssec:da_vs_p}
\begin{figure}
\centering
\begin{minipage}[h]{8.5cm}
\centering
\includegraphics[width=8.5cm]{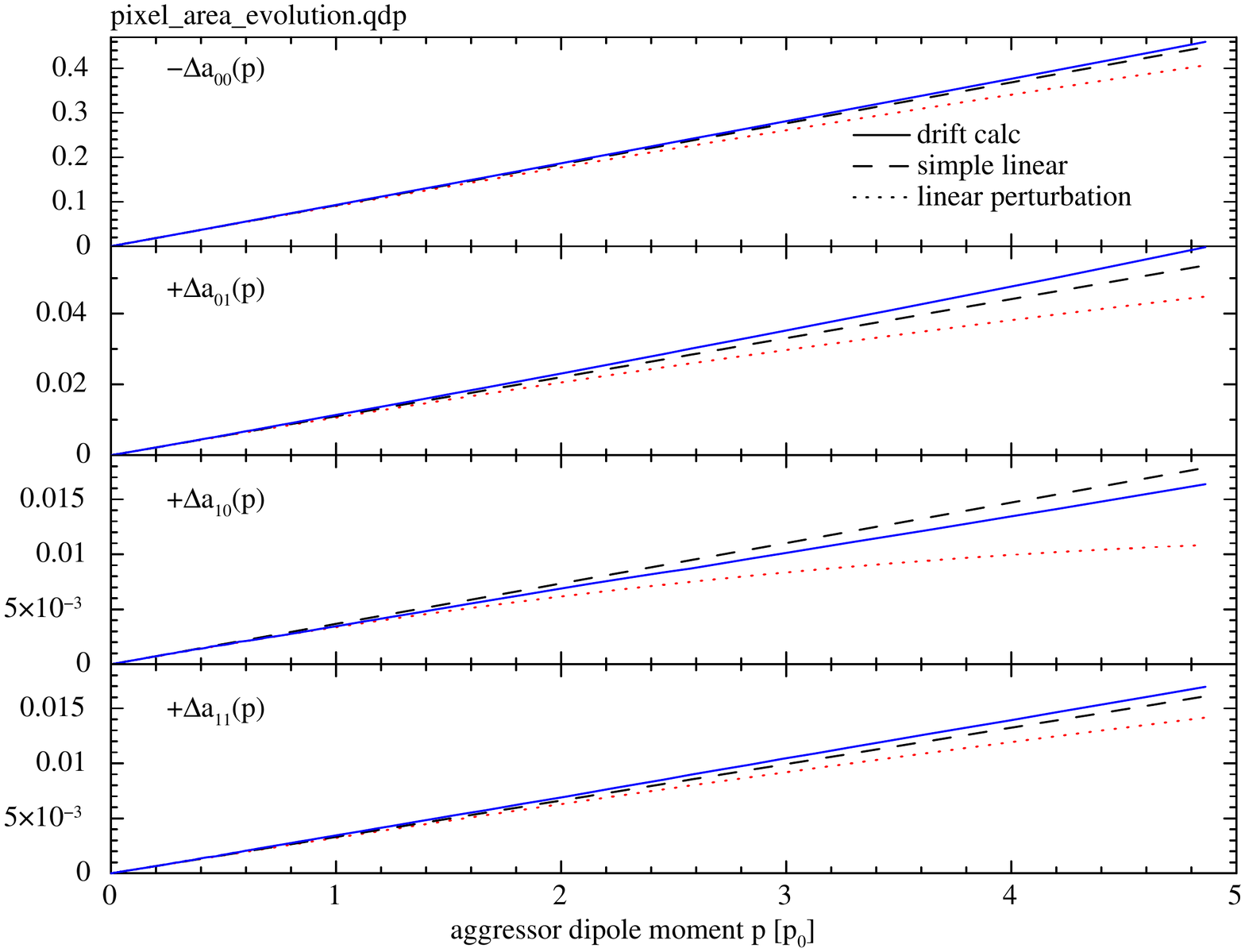}
\end{minipage}
\begin{minipage}[h]{8.5cm}
\centering
\includegraphics[width=8.5cm]{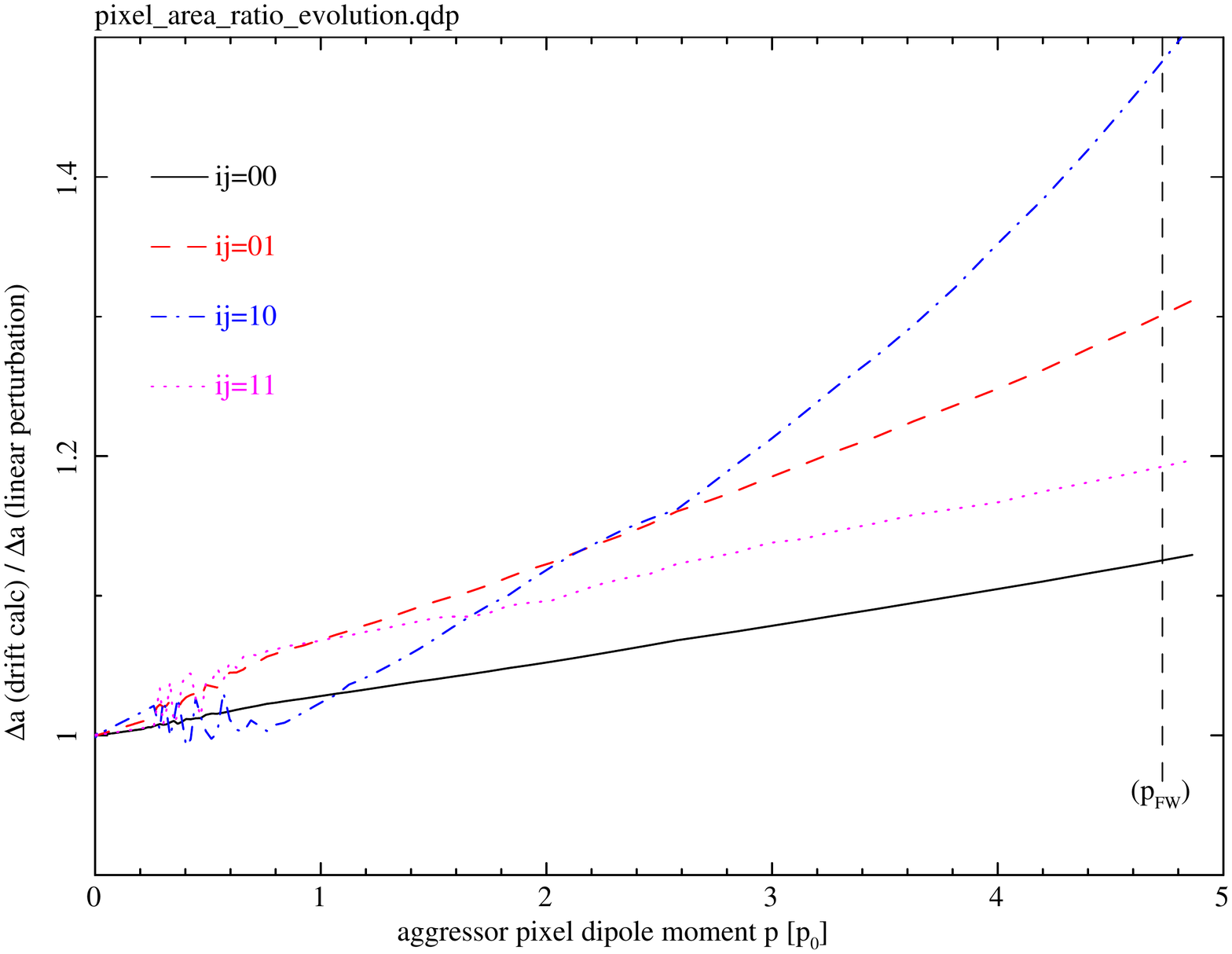}
\end{minipage}
\includegraphics[width=9cm,height=9cm]{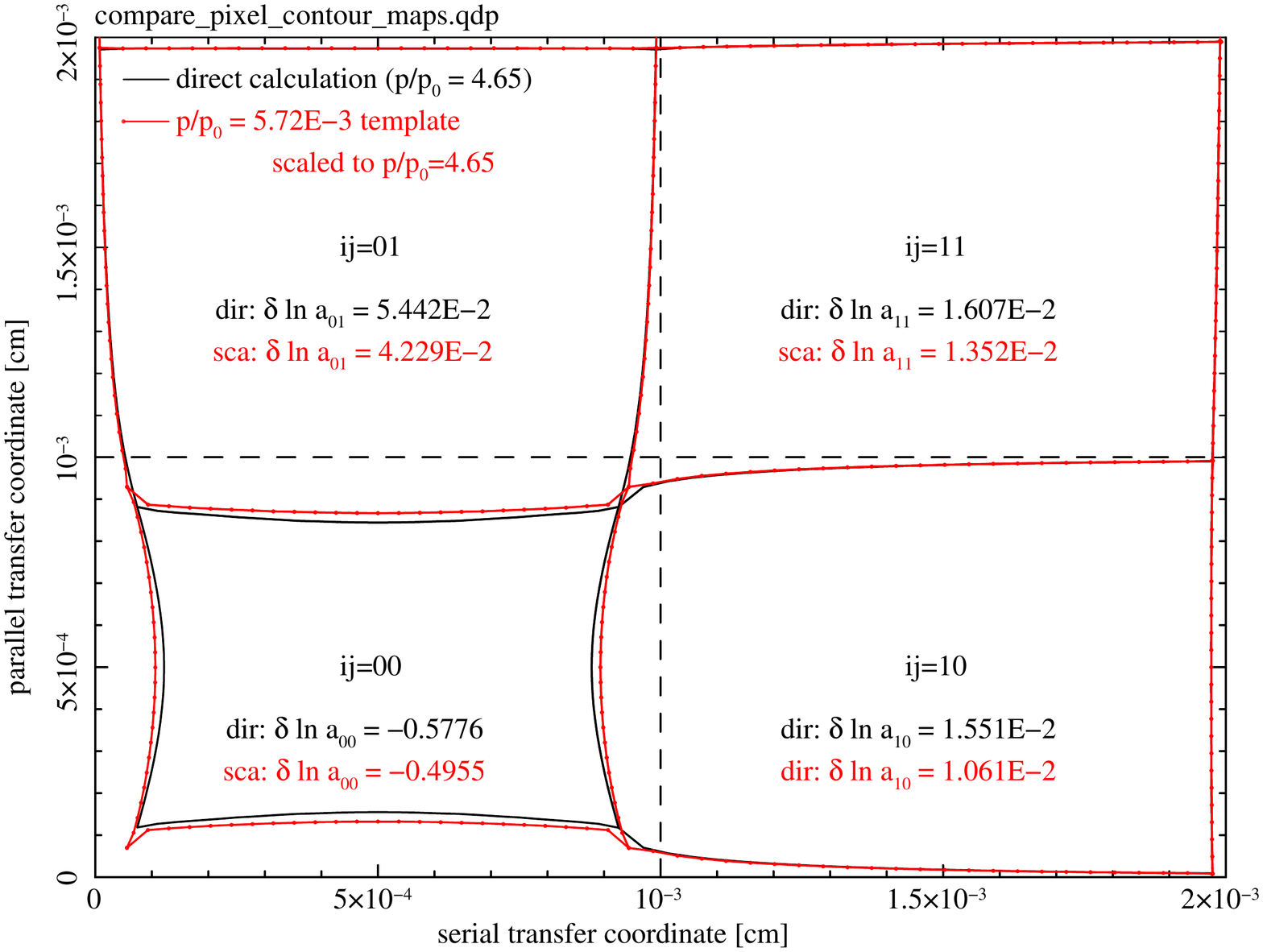}
\caption{
\label{fig:pixel_area_evolution}
A study of pixel area distortion dependence on aggressor amplitude $p$. Upper left: for each of 4 lags, $ij\in \lbrace 00,01,10,11\rbrace$, measures of the pixel area distortions described in the text are given -- the linear perturbation using the {\it BF template} (dotted line), the direct drift calculation (solid line), and a linear area model where the area distortion scales directly with aggressor amplitude $p$ (dashed line). All curves in each plot coincide where $p_t=\bar{p}=0.00572\,p_0$ from the fit to flat field correlations, and are tabulated out to $p\sim 4.7\,p_0$, which corresponds approximately to $100\,\mathrm{ke^-}$ in the aggressor. In the case of $ij=00$ (top tier), a negative sign has been applied to $\Delta a$ for a better comparison to the other curves. Upper right: ratios are plotted to compare the direct drift calculation to the linear perturbation approach using the template for each of the lags considered here. While the two approaches mismatch by about 10\% for $ij=00$ near the full well scale, the discrepancies are typically much larger, as much as 50\%, for the other lags there. Bottom: pixel boundary maps for an aggressor at full well, $100\,\mathrm{ke^-}$ ($p=4.65\,p_0$) to further compare the two methods for these lags. The $\sim$50\% discrepancy between the two methods shown for $ij=10$ may appear counterintuitive because the area gained appears larger than it really is. In fact, $\Delta a_{10} \approx -0.027 \,\Delta a_{00}$, based on the drift calculation shown here.
}
\end{figure}

By evaluating pixel border distortions for different aggressor strengths $p$ and comparing against a linear scaling of the {\it BF template}, we essentially test the validity of the following expressions ({\it cf.} Appendix~\ref{apxsec:template}), where $p_t$ is the aggressor dipole moment for which the template was generated:
\begin{eqnarray}
\delta\vec{c}_k^{\,t(i,j)}(p)&=&\left({p\over p_t}\right) \delta\vec{c}_k^{\,t(i,j)}(p_t),\;\forall k\nonumber\\
\delta d_l^{\,t(i,j)}(p)&=&\left({p\over p_t}\right) \delta d_l^{\,t(i,j)}(p_t),\;\forall l \nonumber
\end{eqnarray}
where the instantaneous $p/p_t$ may be as large as twice the ratio between the canonical full well (100$\mathrm{ke^-}$) to the {\it rms} statistical fluctuation level $\zeta$ given in Table~\ref{tab:bestfit_pars}. In the present case, $p_{max} \sim (2 \mathrm{FW /\zeta})\,\bar{p} \sim 826\,\bar{p}$, many times the fluctuation levels sampled by flat field correlations. Figure~\ref{fig:pixel_area_evolution} provides a quantitative comparison of the evolution of pixel area variations with aggressor $p$ and shows that the scaled template/linear perturbation approach suffers from significant error. It should be emphasized that the apparent (per lag) nonlinearities revealed here are {\it entirely inaccessible} to confirmation via flat field correlations, unless the operational sensor full well is closer to $10^{7}\,\mathrm{ke^-}$: a natural consequence of flat field statistics we used to probe the BF effect in the first place.

% perturbation model using 
% In \S\ref{ssec:da_vs_p}, this notion is tested by comparing against directly computed pixel areas using the same drift model, but for a variable aggressor $p$.
\section{Possible nonlinearities measurable in flat field covariances}\label{sec:ff_nonlinearities}
Although the nonlinearities suggested in \S\ref{ssec:da_vs_p} are not measurable using flat fields for the reasons described, we consider other terms here that could alter parameterization of the electrostatic elements in the drift model, or would otherwise affect charge accumulation in the receiving channels beyond the cold carrier approximation.
\subsection{Aggressor dipole moment dependence on accumulated conversions}\label{ssec:z_twiddle}
The connection between charge collected and effective aggressor dipole moment in this drift model is the depth of the buried channel, $z_{chan}$ ({\it cf.} \S\ref{ssec:application_to_measured_covariances}). A strawman model for the buried channel depth may be constructed by superposing two contributors of the axial component of the drift field, one from the bound charge density (depleted n-type Si) and one from the influence of the free charges in the conductive polysilicon gates, which can be treated as an image charge. With $z_0$ equal to $z_{chan}$ in the limit of an unpopulated channel, we do not allow for any spatial extent to the charge cloud and treat only its centroid:
\begin{eqnarray}
\vec{E}(z_{chan})\cdot\hat{z} &=& \left(\vec{E}^{\,B}(z_{chan}|z_0)+\vec{E}^{\,image}(z_{chan}|z_0)\right)\cdot \hat{z} =0 \nonumber\\
\hat{z}\cdot \vec{E}^{\,B}(z_{chan}|z_0) &=& {\mathrm{N_d} \over 2\epsilon_0\epsilon_{\mathrm{Si}}}\,(z_{chan}-z_0)\nonumber\\
\hat{z}\cdot \vec{E}^{\,image}(z_{chan}|z_0) &=& {\mu + \zeta \over 4\pi\epsilon_0\epsilon_{\mathrm{Si}}(2z_{chan})^2} \nonumber\\
\tilde{z}&\equiv&z_{chan}/z_0\nonumber\\
\tilde{z}^3-\tilde{z}^2&=&-{\mu+\zeta \over8\pi\mathrm{N_d}z_0^3}.\nonumber
\end{eqnarray}
The preceding equations describe two real solutions for finite $\tilde{z}$ (one stable solution for each charge sign) in terms of physical properties of the channel, channel depth for zero signal $z_0$ and the signal occupancy $\mu+\zeta$ (mean plus aggressor). As the channel becomes populated, the  solutions draw closer to one another until they coalesce as an inflection point. Beyond this there is no finite, real solution and the channel should empty as quickly as it fills to expose gate structure layers with conversions. We may define this measure of the full well, $\mathrm{FW} \approx {32\pi\over 27}\mathrm{N_d}z_0^3$ (to potentially constrain physical parameters of the channel using linearity measurements). Solutions for $\tilde{z}$ for $\mu+\zeta \ll {32\pi\over 27}\mathrm{N_d}z_0^3$ follow the approximately linear trend:
\begin{equation}
\tilde{z}-1\approx -{\mu+\zeta\over8\pi\mathrm{N_d}z_0^3} = -{\mu+\zeta\over 3.3\times10^6}\left({\mathrm{N_d}\over{10^{16}\,\mathrm{cm}^3}}\right)^{-1}\left({z_0\over 2.36\,\mu\mathrm{m}}\right)^{-3}.\label{eq:tilde_z}
\end{equation}
If indeed $\mathrm{N_d}\sim 10^{16}\,\mathrm{cm}^{-3}$, then we should expect a weakening trend of the coupling between signal and aggressor dipole moment of about 3\% per $\mathrm{100\,ke^-}$; and a proportionally weakened coupling if $\mathrm{N_d}$ is smaller. This scaling of the coupling term $z_{chan}$ contributes additional, significant detail (depending on $\mathrm{N_d}$) at the  scale shown in Figure~\ref{fig:pixel_area_evolution} and inclusion of this effect is treated in \S\ref{ssec:combined}.

\subsection{Aggressor driven modification to drift time and diffusion contribution to correlations}\label{ssec:sigma_p}
It has been proposed~\cite{Guyonnet_2015,Gruen_jinst_2015} that the presence of collected conversions at the channel can attenuate the axial term in the drift field and could help to explain some of the discrepancies seen between an electrostatic treatment of pixel borders and observed correlations. Guyonnet~\cite{Guyonnet_2015} (\S5.3) investigated this further to place upper limits ($<4\%$) on the contribution by diffusion - longer collection times - to the total the BF effect for focused spots. In the context of the drift model described here, it seems difficult to accurately isolate longer collection times from the effect of distorted pixel boundaries that accompany a reduced electric field near the backside window.  Indeed, an initial survey of temperature dependence of the flat field correlations produced inconclusive results.
In any case, the notion we examine here is that a large aggressor could raise collection times feeding into its own pixel, effectively causing an {\it additional} redistribution of charges to neighboring pixels, while the neighboring pixels do not reciprocate via the same mechanism (they retain their nominal collection times). If significant, this mechanism could be included by adding another term on the right hand side of Equation~\ref{eq:dq_ij}.
\begin{figure}
\begin{minipage}[h]{9cm}
\centering
\includegraphics[width=9cm]{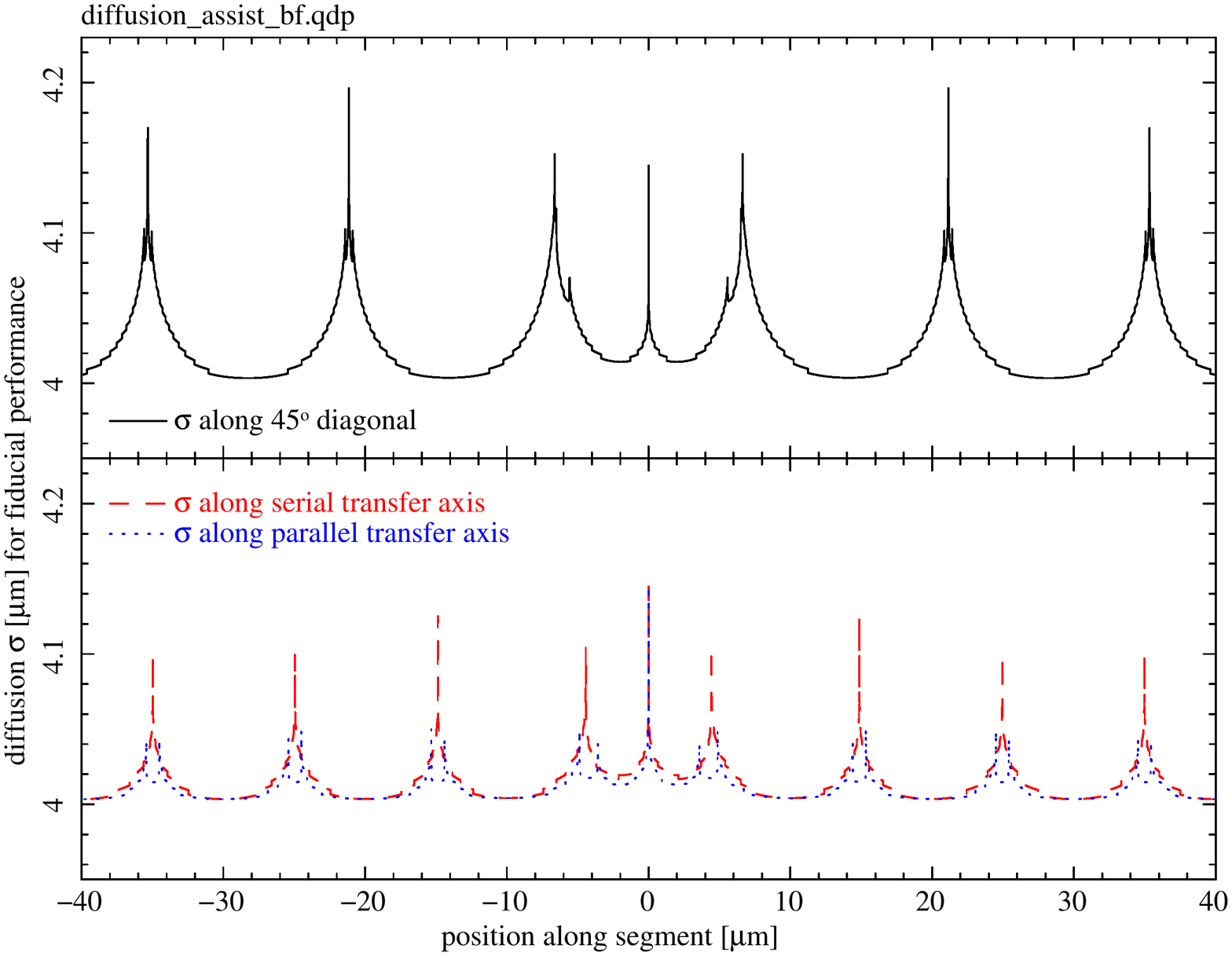}
\end{minipage}
\begin{minipage}[h]{7cm}
\centering
\includegraphics[width=7cm]{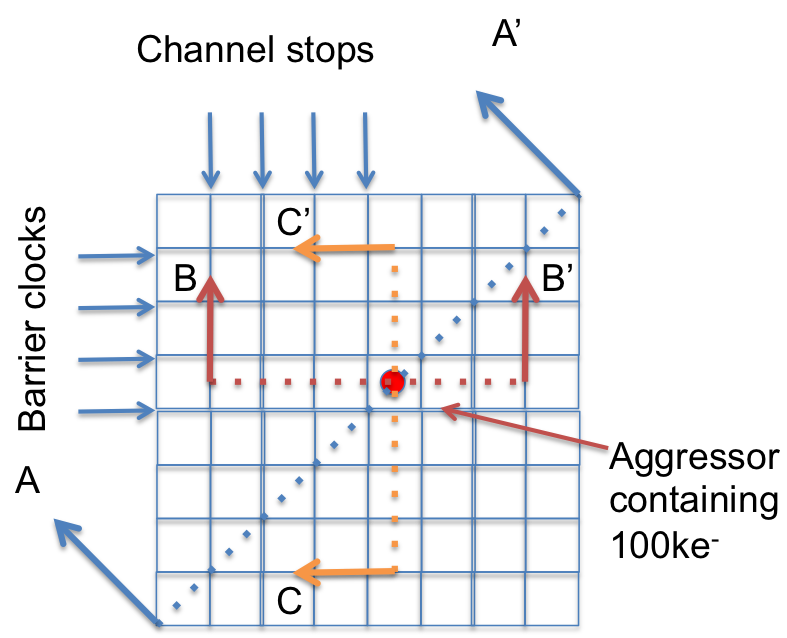}
\end{minipage}
\caption{
\label{fig:diffusion_assist}
A calculation to estimate the {\it diffusion assisted} contribution to the BF effect. Notionally, in addition to the pixel boundaries getting distorted, the presence of the exposure averaged aggressor dipole $\bar{p}$ can slow down the carrier drift toward the channel to increase collection time, boosting the diffusion parameter $\sigma$. For fiducial sensor performance (parameters listed in Table~\ref{tab:bestfit_pars} but also $T=173\,\mathrm{K}$), the drift calculation was used to sample the launch position dependence for diffusion. Left: three linear traces are given -- one diagonal and two along the address axes -- to show $\sigma$ vs. position. 
In each case, position zero is the location of an exposure averaged aggressor dipole $\bar{p}$ that corresponds to a final, recorded signal of $100\,\mathrm{ke^-}$. Presence of the electrostatic elements ($\xi_{cs}$, $\xi_{ck}$ and $\bar{p}$) in specific locations print through to produce the modulations shown. They collectively cause slowdowns and deflections along the field lines. 
Right: a graphic representation of how the linear traces were simulated (sections {\bf A-A'}, {\bf B-B'} and {\bf C-C'}) where lines of the grid represent the barriers that form pixel boundaries. An approximate averaging over these linear traces reveal only a modest (0.25\%) increase in $\sigma$ for positions that feed the central pixel containing the aggressor over the neighbors.
}
\end{figure}

The same drift calculation used to determine pixel boundaries for cold carriers is used to estimate the effect of the accumulated conversions on subsequent conversions' collection times (and diffusion). Figure~\ref{fig:diffusion_assist} shows this dependence as if the trajectory for cold carriers can be used to compute a drift time and diffusion, for carriers with temperature matching the substrate's temperature of $T=173\,\mathrm K$. Presence of the barriers and the aggressor are clearly seen as cusps in the $\sigma(\vec{x})$ field sampled by the crosscuts shown. When averaged over these linear traces, positions nominally tied to the central pixel have a modestly increased $\bar{\sigma}_{00}$ that is about 1 part in 400 greater than for positions not tied to the central pixel. We further estimate the net redistributive effect when the central pixel has a larger diffusion $\sigma$ than its neighbors. With $\sigma_\mathrm{nom}\sim 0.4$ (pixels), we sample and average over all possible Gaussian centroids contained within the pixel to  compute the expected contribution to that pixel, which (for $0.2<\sigma<0.6$) works out to:
\begin{eqnarray}
\left<C_{00}(\sigma)\right>&=&\int_0^1 dx_0 \int_0^1 dy_0 \int_0^1 dx \int_0^1 dy \;{1\over 2\pi\sigma^2} \exp\left(- {(x-x_0)^2+(y-y_0)^2 \over 2\sigma^2} \right) \nonumber \\
&\approx& 1.0216347 - 1.7486435\,\sigma + 0.8989589\,\sigma^2. \nonumber
\end{eqnarray}
The net redistributive effect from lag $ij=00$ to neighboring pixels would be $\delta \ln q_{00} \sim \delta\sigma \frac{\partial}{\partial \sigma}\ln \left<C_{00}\right>$, or about $-2.2\times 10^{-3}$ (for $\sigma\approx 0.4\,p$ and $\delta\sigma \sim \sigma/400$). If this ``missing'' signal were recovered and divided evenly between the nearest four neighboring pixels on average, the largest influence would be seen in lag $ij=10$, because the exposure averaged area increase there ({\it cf.} Figure~\ref{fig:pixel_area_evolution}) is small: $\sim 8\times 10^{-3}$, and would impart a $\sim7\%$, diffusion assisted excess over the nominal, pixel area distortion-driven $\mathrm{Corr}_{10}$. The term is far less consequential for lag $ij=01$ ($\sim2\%$ excess) and for lag $ij=00$ ($\sim 1\%$ excess). 
The fit to the correlations presented above -- Table~\ref{tab:bestfit_pars} and Figure~\ref{fig:pixel_area_model} -- did not include this as a separate term that would tend to dilute the observed anisotropy between $\mathrm{Corr}_{10}$ and $\mathrm{Corr}_{01}$ in the model. It is unknown at this point whether its inclusion would improve the quality of the fit, or if tighter constraints on the $\mathrm{Corr}_{ij}$ measurements would motivate its inclusion. In any case, the term's scaling and dependence mimics that of the area distortion model described in Equation~\ref{eq:dq_ij} and is currently absorbed in the deterministic, detailed pixel boundary model. A separate treatment of this effect may be more important for devices with smaller backside field strengths, or if these devices were operated with smaller $|BSS|$. We are encouraged by what is supported by Figure~\ref{fig:bss_dep} as a reasonable correspondence between measurement and calculation shown for $\mathrm{Corr}_{ij}$ vs. $|BSS|$ -- that separate inclusion of this term may be unwarranted.

\subsection{Predictions for nonlinearities in $\mathrm{\bf Corr}_{\bf ij}{\bf ( \mu)}$}\label{ssec:combined}

In the sections above, several terms were described that affect the evolution of the dynamic pixel area distortion model that are currently not well constrained by test stand measurements. These generally influence the detailed relationship between accumulated flux and aggressor dipole moment (\S\ref{ssec:z_twiddle}). In the current case, we have a pattern of $\mathrm{Corr}_{ij}$ measured at a single flat field level $\mu$ and variance $\zeta^2$, a drift model parameter list that reasonably reproduces the correlation pattern when pixel area distortions are mapped via Equation~\ref{eq:corij}, and a set of pixel boundary distortions generated for a selection of aggressor levels $p$. Assuming for the time being that the gain was accurately determined and that errors on $\mu$ and $\zeta$ are negligible, we investigate how the computed area distortion patterns can be mapped onto an observable set of parameters.  By first assuming a value for $\mathrm{N_d}$, expressions for the aggressor $\bar{p}$ and $\tilde{z}$ are used to determine channel depth in the zero signal limit by solving iteratively
\begin{eqnarray}
z_0 &=& {\bar{p}_F / \zeta_F \over 1- {\mu_F \over 16 \pi \mathrm{N_d}z_0^3}},\nonumber
\end{eqnarray}
where $\mu_F$ and $\zeta_F$ were used in the fitting procedure (Eq.~\ref{eq:corij}) used to determine $\bar{p}_F$. Drift calculations for other aggressors $\bar{p}_k$ produce the pixel area distortions $\Delta a_{ij}(\bar{p}_k)$. Self-consistent mean \& variance pairs ($\mu_k^{\scriptstyle{'}},\zeta_k^{\scriptstyle{'}}$) are then calculated, also iteratively, using the equations:
\begin{eqnarray}
\zeta_k^{\scriptstyle{'}} &=& {\bar{p}_k/z_0 \over \left(1 - {\mu_k^{\scriptstyle{'}}\over 16\pi\mathrm{N_d}z_0^3}\right)} = {A\over1-B\mu_k^{\scriptstyle{'}}};\nonumber \\
\mu_k^{\scriptstyle{'}}&=& \left(\zeta_k^{\scriptstyle{'}}\right)^2 \left(1-{\mu_k^{\scriptstyle{'}} \over \zeta_k^{\scriptstyle{'}}}\Delta a_{00}(\bar{p}_k){\left(1-{\mu_k^{\scriptstyle{'}} \over 16\pi\mathrm{N_d}z_0^3}\right)\over\left(1-{\mu_F \over 16\pi\mathrm{N_d}z_0^3}\right)}\right) = 
\left(\zeta_k^{\scriptstyle{'}}\right)^2 \left(1-{\mu_k^{\scriptstyle{'}} \over \zeta_k^{\scriptstyle{'}}}\Delta a_{00}(\bar{p}_k){1-B \mu_k^{\scriptstyle{'}} \over 1-B\mu_F}\right). \nonumber
\end{eqnarray}
This process allows us to predict the detailed shape of the mean-variance curve as well as the mean-correlation curves for specific lags $ij$. A similar procedure would be used to allow for (and constrain) a gain error in a non-degenerate way. This isn't discussed here, but is straightforward to implement, given an additional set of approximate, ($\mu$,$\zeta$) pairs derived from photon transfer curves. 

\begin{figure}
\centering
\begin{minipage}[h]{8.5cm}
\centering
\includegraphics[width=8.5cm]{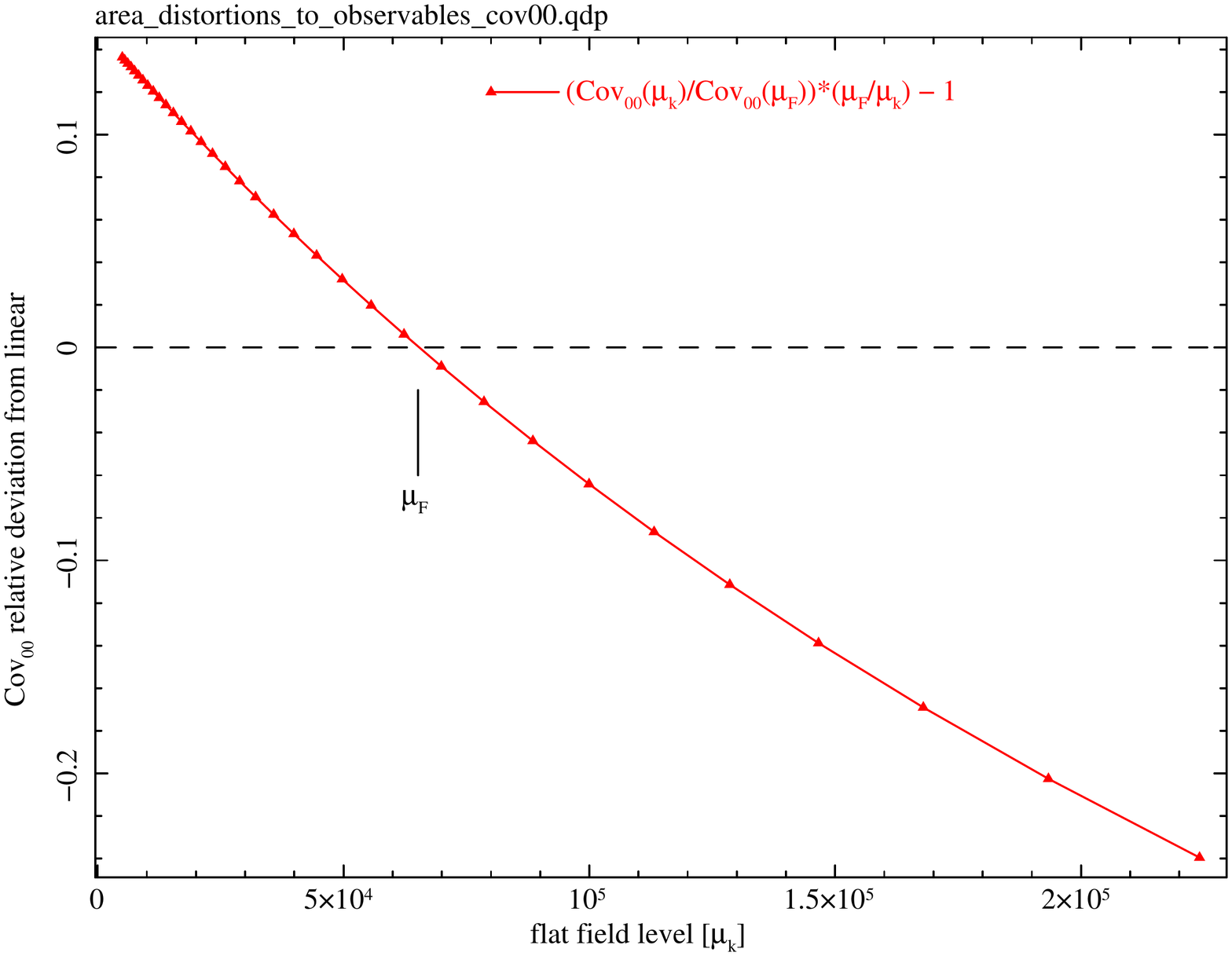}
\end{minipage}
\begin{minipage}[h]{8.5cm}
\centering
\includegraphics[width=8.5cm]{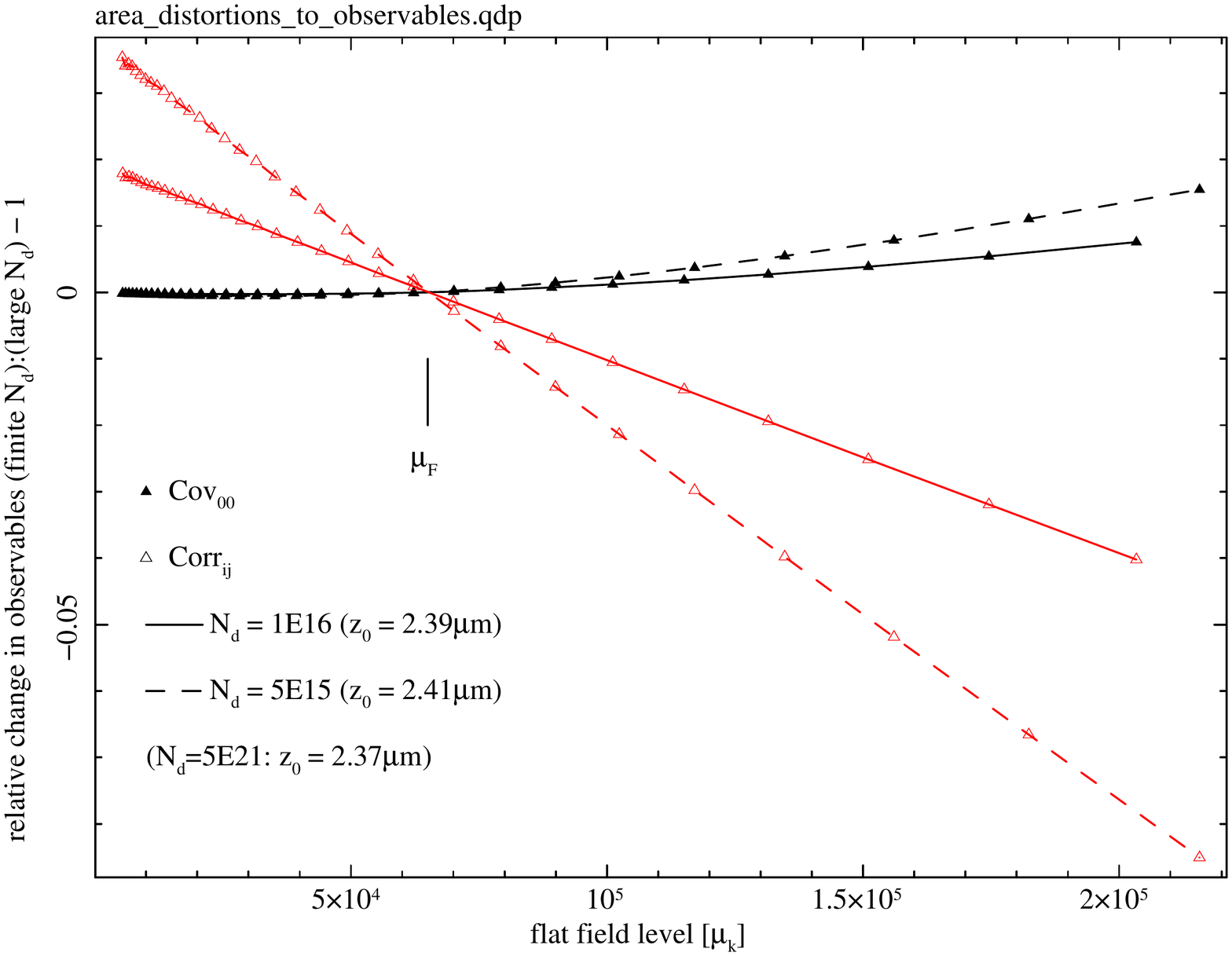}
\end{minipage}
\caption{
\label{fig:nlo_in_observables}
Aggressor induced, pixel area distortion calculations ($\Delta a_{ij}(\bar{p}_k)$) mapped into observable flat field statistics to predict their dependence on flat field level $\mu_k$, while allowing for an unknown donor density in the channel ($\mathrm{N_d}$). Left: the expected nonlinear term in the variance, $\zeta_k^2 \equiv \mathrm{Cov}_{00}(\mu_k)$, evaluated for a fixed channel depth (equivalently, a large $\mathrm{N_d}$). Any existing error in the gain determination should result in photon transfer data plotted off of the locus shown here. The level called out as $\mu_F$ is the flux level at which the correlation pattern was fit with an adequate pixel area distortion template drawn using the drift calculation. Similar nonlinear terms in calculations of $\mathrm{Corr}_{ij}$ were below the percent level with no clear trend, so these were not shown here. Right: corrections to the $\mathrm{Cov}_{00}$ and $\mathrm{Corr}_{ij}$ mappings that result from a channel depth that evolves with signal level. Two values for donor density in the channel are shown: $\mathrm{N_d}=1\times10^{16}$ and $5\times10^{15}\,\mathrm{cm^{-3}}$. Mapping corrections for $\mathrm{Cov}_{00}$ tend to be limited to the 2\% level, but corrections to $\mathrm{Corr}_{ij}$ are independent of lag $ij$, and would tend to show small positive offsets in $\mathrm{Corr}_{ij}$ vs. $\mu$, depending on the sampling values that are available.
}
\end{figure}

Figure~\ref{fig:nlo_in_observables} provides a family of curves that predict the signal dependence of the observable quantities $\zeta^2\equiv \mathrm{Var}$ and $\mathrm{Corr}_{ij}$ for different assumptions of $\mathrm{N_d}$, again by assuming that an accurate gain was determined to produce $\mu_F$ and $\zeta_F$. Self-consistent values for $z_0$ for the assumed values of $\mathrm{N_d}$
are also given. It turns out that while the mean-variance curve is relatively insensitive to the nonlinearities considered in \S3.3, the signal level dependence of the $\mathrm{Corr}_{ij}$ may the most straightforward indicator for an evolution in the coupling between signal and aggressor.

\section{Falsifiable tests of linear extrapolations of the brighter-fatter template: high-contrast laboratory tests}\label{sec:hi_contrast}

The situation we described above is that we predict significant departures from linear perturbation models when we deal with real, high-contrast/high dynamic range data. Difficulty arises from not being able to sample high-contrast/high dynamic range conditions using flat fields alone: the aggressor scale available tops out near the square root of the full well. It's surprising, then, that the linear perturbation methods used in astronomical pixel data pipelines can correct 90\% of these dynamical effects: that only 10\% of the initial BF effect remains uncorrected\cite{Lupton:privcomm} after the compensation strategy is applied. This is based only on tracking a single ``width'' parameter for the PSF's intensity dependence, and does not at all capture the platykurtic distortions to the PSF profile that result from the boundary distortion mechanism. Nonlinear terms due to the variable channel depth appear to reduce the BF effect by about 6\% averaged over the exposure for full well (if $\mathrm{N_d}=5\times10^{15}\,\mathrm{cm^{-3}}$); the direct drift calculations suggest that the linear perturbation underestimates the BF effect, anisotropically, by $10-20\%$ for high-contrast/high dynamic range exposures reaching the same full well in the pixels receiving the highest flux. We expect that sensors using smaller electric fields or longer drift distances than these LSST candidate devices should show correspondingly larger complications.

\begin{figure}
\centering
\begin{minipage}{8cm}
\centering\includegraphics[width=8cm]{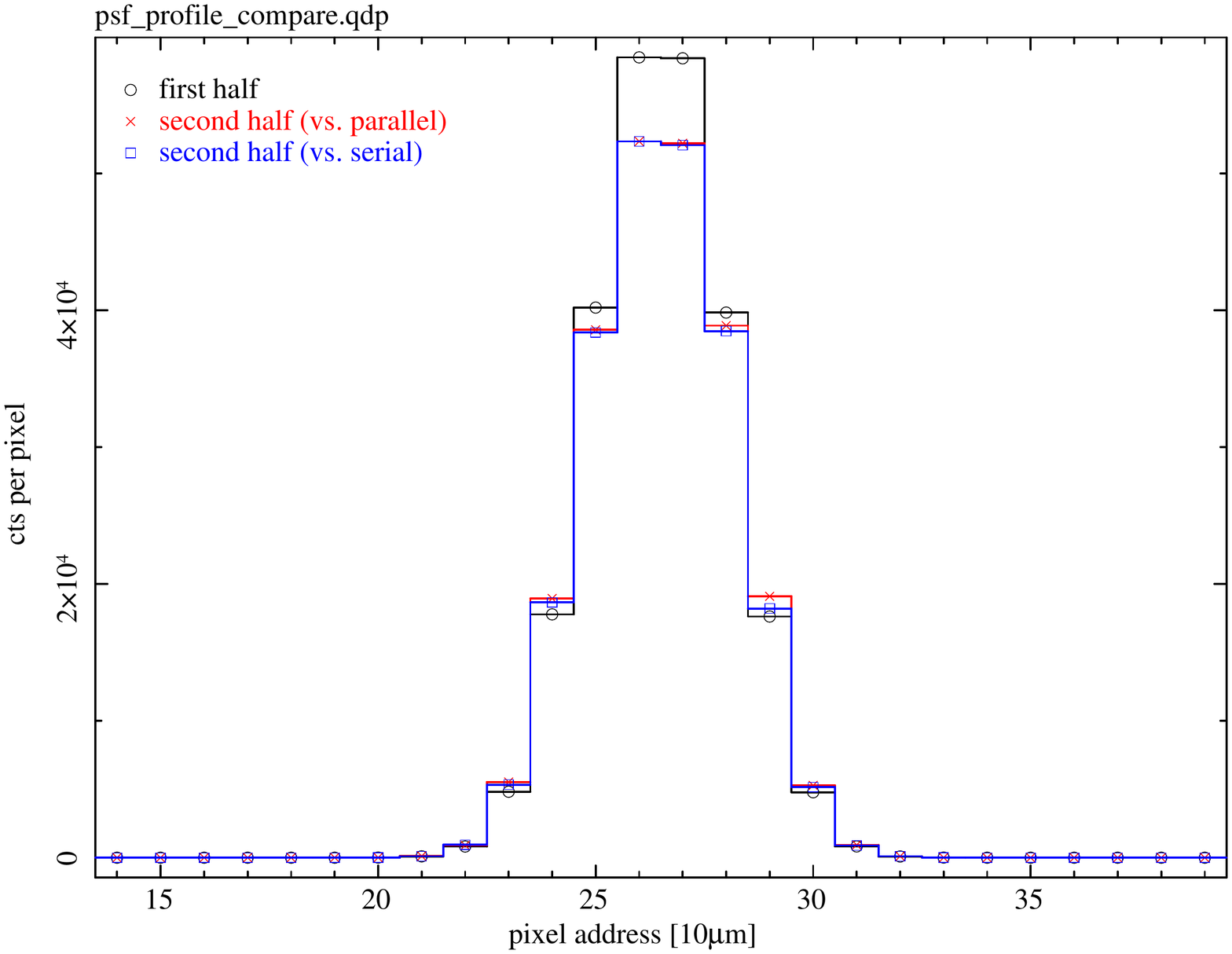}
\end{minipage}
\begin{minipage}{8cm}
\centering\includegraphics[width=8cm]{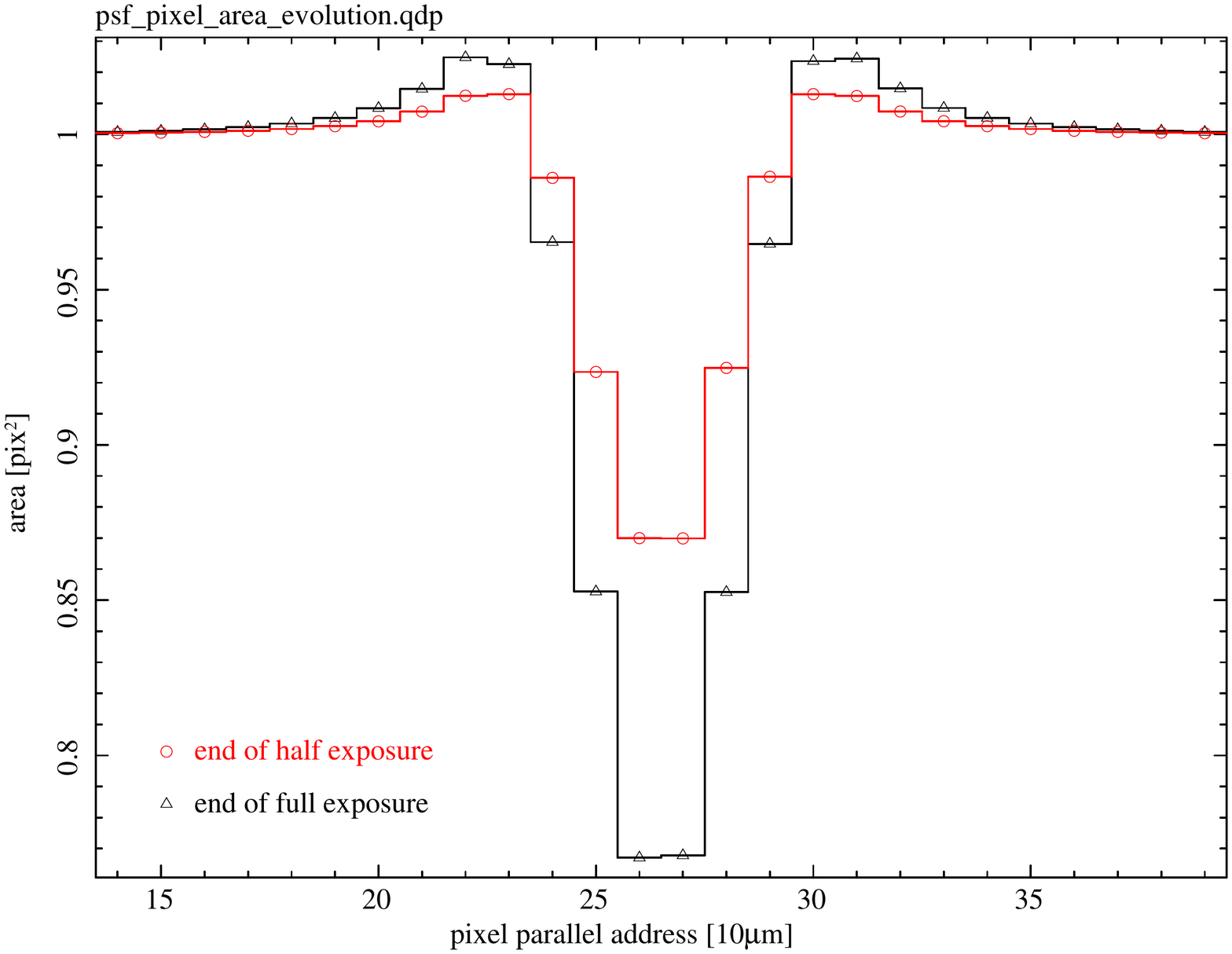}
\end{minipage}
\caption{
\label{fig:psf_evolution}
A high-contrast, high dynamic range simulation of an idealized PSF using linear perturbation of the template. The illumination used is an isotropic Gaussian with 0.7$^{\scriptstyle{''}}$ FWHM centered on the geometric midpoint of four pixels. Integration continues until full well is reached, which corresponds to AB$\sim$15.2 for LSST's {\it r}-band in a 15 second exposure. Left: a comparison of the accumulated image after the first half of the exposure to the additional accumulation during the second half. These are linear traces that pass through the PSF centroid. The BF effect is seen by comparing linear traces for the second half of the integration against that for the first half. The total number of counts in the traces for the second half are typically about 5\% lower than for the first half, because counts are also distributed perpendicularly to the trace. Right: A comparison of the pixel areas resulting from the PSF integration half-way through and after completion. These can be used to estimate structure in the sky background contribution, and errors in the PSF profile if sky background is subtracted (without using this information). Laboratory data obtained to reproduce these results may reveal the differences predicted in \S\ref{ssec:combined} and the limitations of the linear perturbation model used here.
}
\end{figure}

\begin{figure}
\centering
\begin{minipage}{3cm}
\centering\includegraphics[width=3cm]{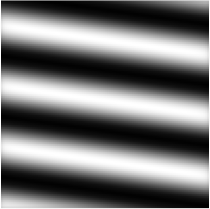}
\end{minipage}
\begin{minipage}{12cm}
\centering\includegraphics[width=12cm]{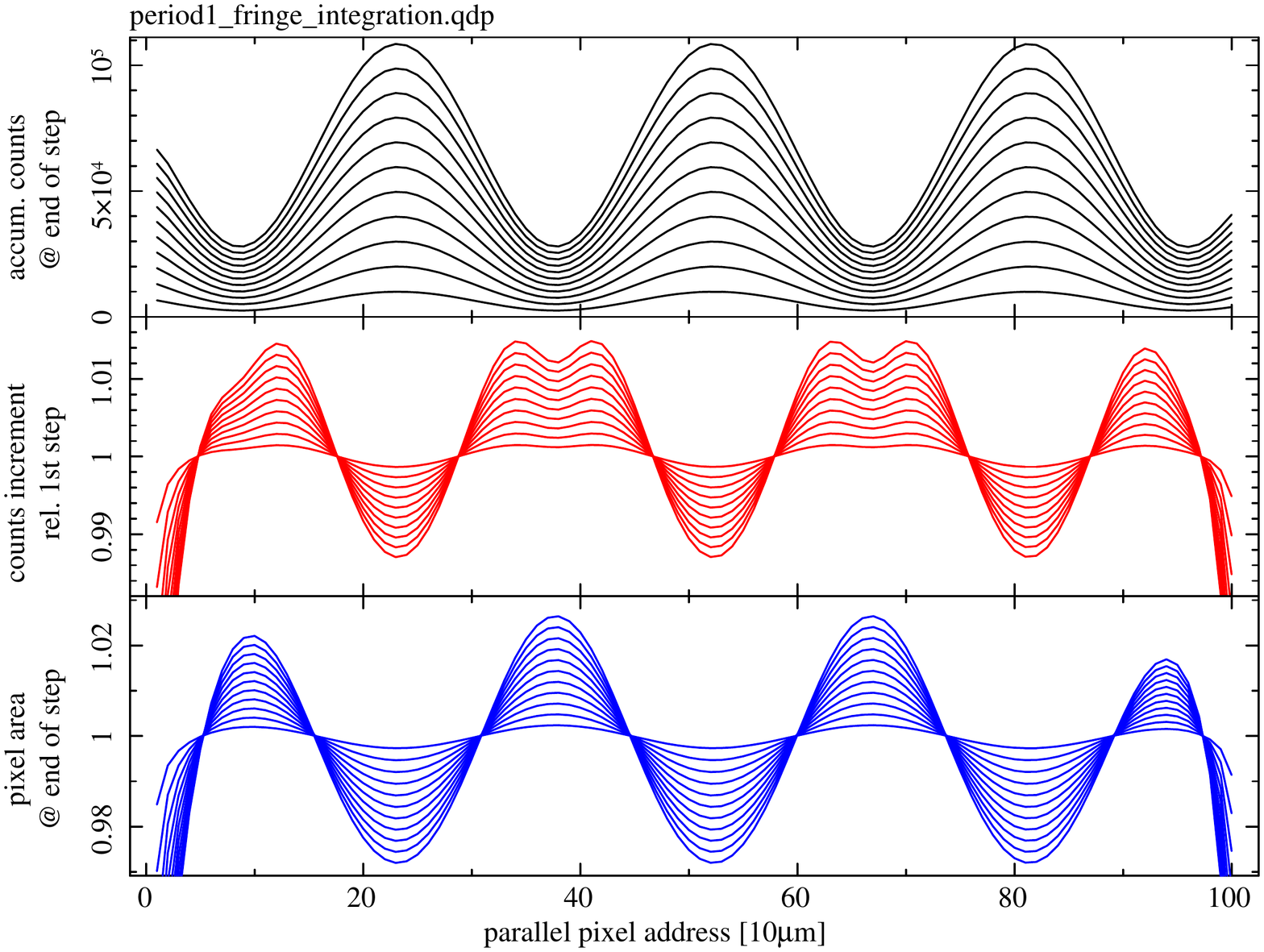}
\end{minipage}\\
\begin{minipage}{3cm}
\centering\includegraphics[width=3cm]{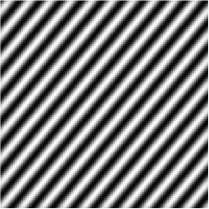}
\end{minipage}
\begin{minipage}{12cm}
\centering\includegraphics[width=12cm]{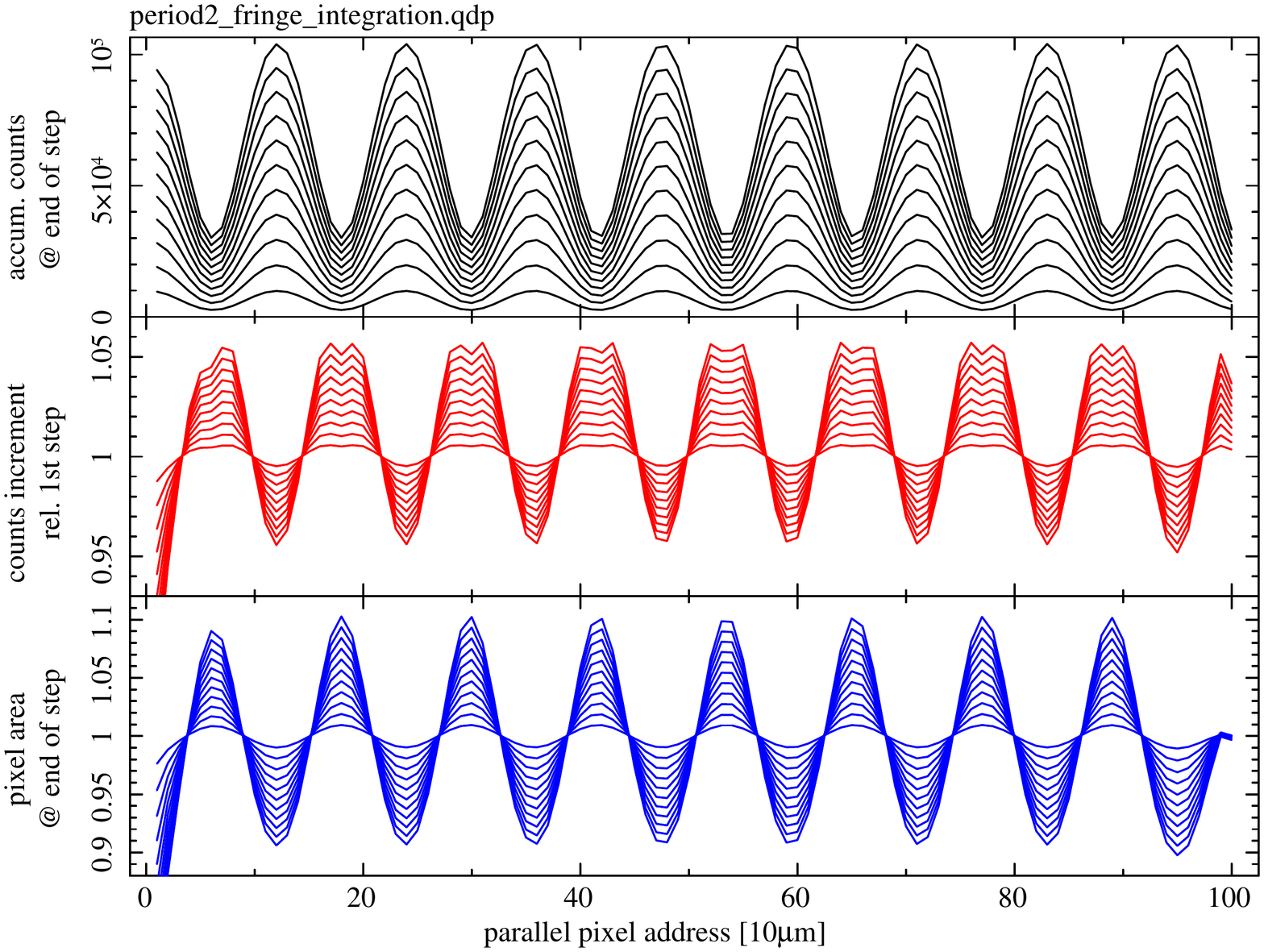}
\end{minipage}
\caption{
\label{fig:fringe_evolution}
A high-contrast, high dynamic range simulation of fringe projector illumination to accompany laboratory measurements. This simulation used linear perturbation of the template and the illumination was chosen to imitate the available parameter space. In both cases, the peak:valley ratio is set to 3, and the final maximum counts accumulated is near the full well depth of $100\,\mathrm{ke^-/pixel}$. Postage stamp images to the left show the image accumulated at readout time. Upper plots: a fringe period of 28.8 pixels with orientation 82$^\circ$; Lower plots: a fringe period of 7.6 pixels with orientation 140$^\circ$. On the right, for each fringe calculation, the plots show the counts accumulated in 11 steps (top tiers), the 10 ratios of the incremental counts accumulated divided by counts accumulated in the first step (middle tiers), and the 11 area curves computed at the end of each step (bottom tiers). The perturbed area fields may be validated by subtracting ``fringe only'' exposures from ``fringe+flat'' exposures, which may be the most direct way to probe a dynamic pixel distortion mechanism. The perturbed area field would be easier to measure, with greater contrast, if the ``flat'' part of the exposure could be applied {\it after} the ``fringe'' part, rather than in a simultaneous exposure.
}
\end{figure}

We consider some lab measurements that could be performed to test the drift model -- and the linear perturbation template methodology. Because the template is based only on a single aggressor level $\hat{p}_F/p_0=0.00572$ (for $\mu_F=65\,\mathrm{ke^-}$ \& $\zeta_F=242\,\mathrm{e^-}$), the next-to-leading order terms described in \S\ref{ssec:combined} are not carried. We describe a receiving pixel array with geometric parameters that evolve with time, as the exposure progresses toward full well. Ratios of images (long vs. short exposure), incremental difference images (subtraction of images with adjacent exposure times), etc., are simulated. Pixel areas at the end of exposure are also recorded -- to predict the effect of a flat field ``flash'' at the end of the high-contrast exposure. Differences between (high-contrast + flash) and (high-contrast only) can be used to measure the pixel area field across the array at the end of high-contrast exposure. Deviations from these predictions may be interpreted as a superposition of the nonlinearities described, namely details of the lag-dependent pixel area evolution with aggressor, combined with details of the relationship between signal and aggressor, and the evolution of the channel depth. Figures~\ref{fig:psf_evolution} and \ref{fig:fringe_evolution} show predictions for a focused, Gaussian spot and for interferometric two-slit fringe projections, respectively. NB: these predictions do not include the detailed, position-dependent drift times for cold carriers (\S\ref{ssec:sigma_p}) and only use the one-to-one position-to-pixel mapping. 

% \subsection{Evolution of a quiescent PSF with integrated counts}\label{ssec:psf_evolution}

%\subsection{Background distribution in the presence of pixel area distortions}\label{ssec:bkg_dist}
%\subsection{Fringe projector simulations}\label{ssec:fringe}

\section{Conclusions}
By mapping and matching drift calculation results, representing aggressor-victim pixel area distortions, to measured flat field correlations, we whittled down and simplified the {\it BF effect} to just three electrostatic parameters of the drift field that were not already well constrained by X-ray charge cloud diffusion, for this CCD imaging sensor. A fourth, self-interaction parameter was also determined, and defines the efficiency by which conversions collected in the channel can break the symmetry of the drift field to reposition and distort pixel boundaries encountered for subsequent conversions. 

A compact, digital form for the pixel boundary mapping kernel, or Green's function, was used to predict results for some high-contrast, high dynamic range illuminations of the sensor that could be tested in the laboratory. We expect measurable deviations from these predictions for at least two reasons: (1) that the pixel area distortion variation with aggressor amplitude does not increase as the linear, perturbation theory would predict, and (2) that the self-interaction parameter coupling should evolve with accumulated conversions as an exposure progresses. 

These laboratory measurements, when performed, may provide a basis by which our quantitative understanding of the BF mechanism can be extended to include the high-contrast, high dynamic range domain needed for precision astronomical corrections.

% Appendices

\appendix

\section{Comparisons to similar expressions in the literature}\label{apxsec:expressions}

Because we aim to generalize and extend what's already in the literature, it should be useful to review here expressions of similar quantities that have already been published. Note that in the preceding equations, $a_{ij}$ indicates pixel areas [{\it e.g.,} cm$^2$], {\it not} pixel boundary shift coefficients perpendicular to boundary axes [{\it e.g.,} pixel/carrier], as in some of the expressions below.

\subsubsection{{Antilogus {\it et al.} (2014)}}
In their \S4.2 treatment of charge responsive pixel boundaries applied to flat field correlations, the authors don't distinguish between {\it instantaneous} pixel boundary shifts and those shifts implied by statistics of the {\it recorded image} -- in other words, the {\it exposure averaged} boundary shifts. We find a factor of 2 discrepancy between their equations and ours if the former interpretation is followed, but perhaps no discrepancy with the latter. Upon comparing their equations~4.14 and 4.15 against our approximate expressions (Eqs.~\ref{eq:covij_approxb} and \ref{eq:cov00b} respectively):
\begin{align*}
\mathrm{Cov}\left(Q_{i,j}^{\scriptstyle '}, Q_{0,0}^{\scriptstyle '}\right) &= 4\mu V \sum_X a_{i,j}^X \\
% \rightarrow \mu V {d \ln a_{ij} \over dq_{00}} = 2\,\mathrm{Cov}_{ij}.
\label{eq:antilogus_cov00}
\mathrm{Cov}_{ij} &\approx {\mu\over 2}\,\mathrm{Cov}_{00}\,{d\ln a_{ij} \over dq_{00}}
\end{align*}
and
\begin{align*}
\mathrm{Cov}\left(Q_{0,0}^{\scriptstyle{'}},Q_{0,0}^{\scriptstyle{'}}\right)&=V+4V\mu\sum_X a_{0,0}^X\\
\mathrm{Cov}_{00} &\approx {\mu\over 1-{\mu\over 2}{d\ln a_{00}\over dq_{00}}} \approx \mu \left(1  + {\mu\over 2}{d\ln a_{00} \over dq_{00}} \right) + O\left(\mu^3 \left({1 \over 2} {d\ln a_{00} \over dq_{00}}\right)^2\right).
\end{align*}
We identify the equivalent {\it instantaneous} area distortion coefficients $d\ln a_{kl}/dq_{00}$ that are a factor of 2 larger than the exposure averaged area distortion coefficients, valid for flat field applications, at least:
\begin{align*}
{1\over 2} {d\ln a_{ij}\over dq_{00}} &\equiv {d\ln \bar{a}_{ij}\over dq_{00}} \sim {V^{Eq.4.14} \over \mathrm{Cov}_{00}} \;4 \sum_X a_{i,j}^X\\
{1\over 2} {d\ln a_{00}\over dq_{00}} &\equiv {d\ln \bar{a}_{00}\over dq_{00}} \sim {V^{Eq.4.15}\over\mu} \;4 \sum_X a_{0,0}^X+O\left(\mu\left({1\over 2}{d\ln a_{00}\over dq_{00}}\right)^2\right)
\end{align*}
where it appears that the $V$ in their equations~4.14 and 4.15 may have different definitions.\footnote{If indeed $V^{Eq.4.14}=V^{Eq.4.15}$, then Equations~4.14 and 4.15 can be used together with Eq.~4.4, the sum rule for $a_{i,j}^X$, to recover Poisson statistics, essentially by rebinning. Coefficients to the $a_{i,j}^X$ terms cancel, leaving $\mu=V$. However, we believe Equation~4.14 should scale with the recorded variance: $V^{Eq.4.14}\neq\mu$.
% Equation~4.14 connects the three quantities $\mathrm{Cov}(Q_{i,j}^{\scriptstyle '},Q_{0,0}^{\scriptstyle '})$, $\mu$ and $V$, where $V$ is presumably the measured variance, {\it AKA} $\mathrm{Cov}(Q_{0,0}^{\scriptstyle '},Q_{0,0}^{\scriptstyle '})$. Meanwhile, Equation~4.15 connects $\mathrm{Cov}(Q_{0,0}^{\scriptstyle '},Q_{0,0}^{\scriptstyle '})$, $\mu$ {\it and} $V$: with $V$ described as the {\it input variance} -- presumably this is equal to $\mu$ according to Poisson statistics if no dynamic boundary shifts occur.
}

\subsubsection{Guyonnet {\it et al.} (2015)}
In their \S5.2 parameterization of pixel size variations as a function of flux, this paper uses largely the same notation as in \cite{Antilogus_paccd_2014}, except that boundary shift coefficients $a_{i,j}^X$ are defined differently by a factor of 4, such that the area distortion coefficients are identified, but the definitions for their $V$ in equations~16 and 17 remain to be aligned:\footnote{Augustin notes that high-quality fits to covariances for $(i,j)\neq (0,0)$ are achieved if the following relation is assumed: $2V^{Eq.16}=\mu+\mathrm{Cov}\left(Q_{0,0}^{\scriptstyle '},Q_{0,0}^{\scriptstyle '}\right)$}
\begin{align*}
{1\over 2} {d\ln a_{ij}\over dq_{00}} &\equiv {d\ln \bar{a}_{ij}\over dq_{00}} \sim {V^{Eq.16} \over \mathrm{Cov}_{00}} \;\sum_X a_{i,j}^X\\
{1\over 2} {d\ln a_{00}\over dq_{00}} &\equiv {d\ln \bar{a}_{00}\over dq_{00}} \sim {V^{Eq.17}\over\mu} \;\sum_X a_{0,0}^X+O\left(\mu\left({1\over 2}{d\ln a_{00}\over dq_{00}}\right)^2\right).
\end{align*}

\subsubsection{Gruen {\it et al.} (2015)}
In their \S3.2 (Flat field covariances), the authors cite~\citenum{Antilogus_paccd_2014} but yet write down slightly different expressions for the covariances as a function of lag, and choose a different normalization for the pixel boundary shifts. As before, we compare their Equations~3.7 and 3.8 to approximate our approximate forms of our Eqs.~\ref{eq:covij_approxb} and \ref{eq:cov00b}, respectively:
\begin{align*}
\mathrm{Cov}\left(Q_{00},Q_{ij}\right)&=2\mu^2\sum_{X=T,B,L,R}a_{ij}^X\\
Cov_{ij}&\approx {\mu\over 2}\mathrm{Cov}_{00}{d\ln a_{ij}\over dq_{00}}
\end{align*}
and
\begin{align*}
\Delta \mathrm{Var}\left(Q_{00}\right)&=\mathrm{Var}-\mu = -4\mu^2\left(a_{1,0}^R+a_{0,1}^T\right) = +2\mu^2\sum_{X=T,B,L,R}a_{00}^X\\
\mathrm{Cov}_{00}-\mu&\approx {1\over 2}\mu^2{d\ln a_{00}\over dq_{00}}+O\left(\mu^3\left({1\over 2}{d\ln a_{00}\over dq_{00}}\right)^2\right).
\end{align*}
Equivalent instantaneous area distortion coefficients $d\ln a_{kl}/dq_{00}$ are expressed as:
\begin{align*}
{1\over 2} {d\ln a_{ij}\over dq_{00}} &\equiv {d\ln \bar{a}_{ij}\over dq_{00}} \sim {\mu^{Eq.3.7} \over \mathrm{Cov}_{00}} \;2 \sum_X a_{i,j}^X\\
{1\over 2} {d\ln a_{00}\over dq_{00}} &\equiv {d\ln \bar{a}_{00}\over dq_{00}} \sim 2 \sum_X a_{0,0}^X+O\left(\mu\left({1\over 2}{d\ln a_{00}\over dq_{00}}\right)^2\right)
\end{align*}
where the authors have assumed $\mathrm{Var}=\mu$ in the flat image prior to expressing the change in $\mathrm{Var}$ in Eq.~3.8.

It should be pointed out that all of the above sets of expressions each internally recover Poisson statistics as covariances out to large lags $ij$ are summed: The equations of \cite{Antilogus_paccd_2014,Guyonnet_2015} and \cite{Gruen_jinst_2015} have terms that cancel by subtraction, while (our) equations~\ref{eq:covij_approxa} and \ref{eq:cov00a} are derived from this same principle. Also, the mean--variance relation out to large signal levels $\mu$ may be calculated recursively if $\Delta\ln a_{ij}\left(\mu+\zeta|\mu\right)$ (Eq.~\ref{eq:approx}) is computable.

\section{Electrostatic drift model for cold electrons}\label{apxsec:estat_summary}
\begin{figure} [ht]
\begin{center}
\begin{tabular}{c} %% tabular useful for creating an array of images 
\includegraphics[width=0.7\linewidth]{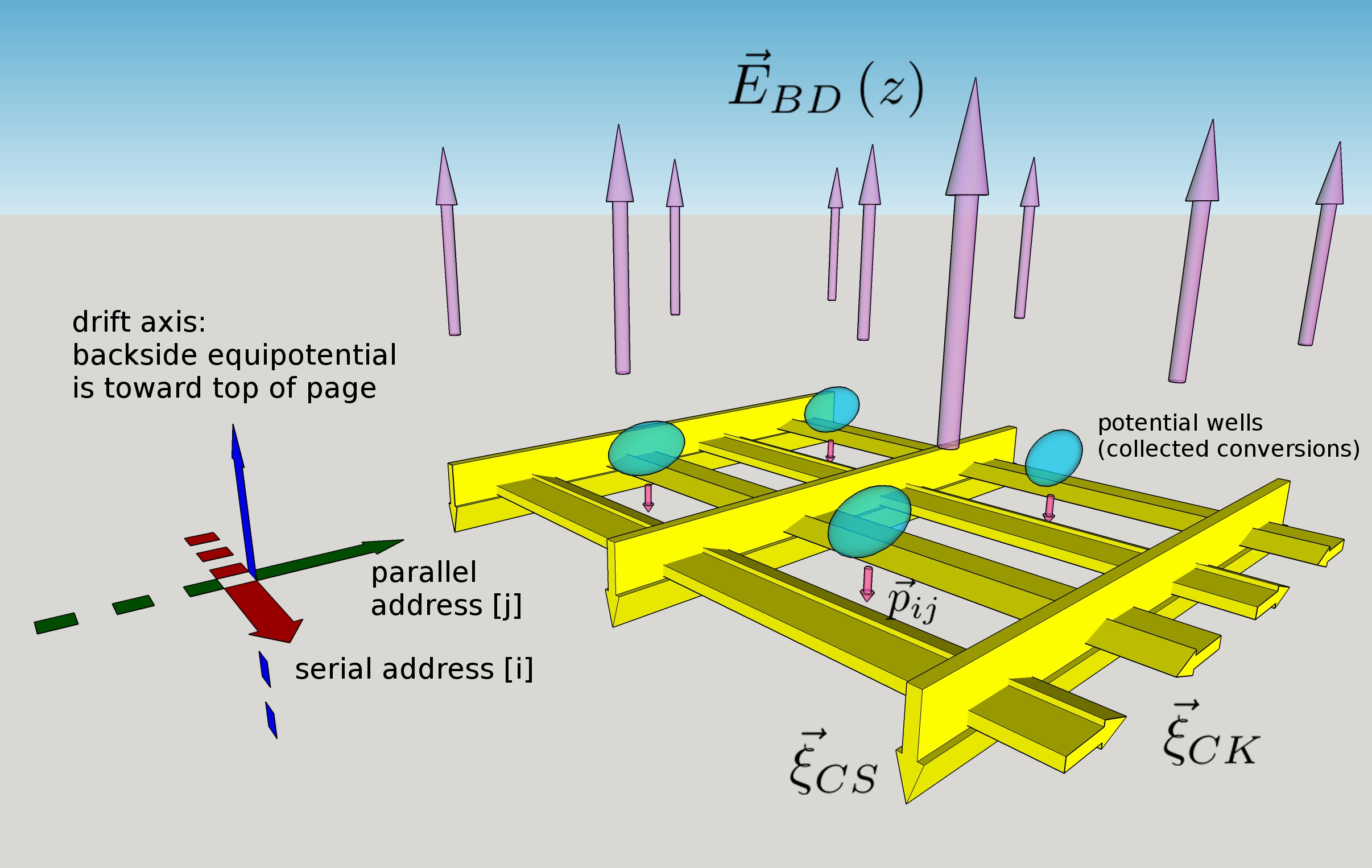}\\
\end{tabular}
\end{center}
\caption{
\label{apxfig:sensor_drift_geometry} 
The assumed field geometry in the electrostatic model.
}
\end{figure}

Here we briefly summarize the electrostatic drift field calculation, which is described in greater detail elsewhere\cite{Rasmussen_paccd_2015,Rasmussen_spie_2014,Rasmussen_paccd_2014} and reproduced here for convenience. Figure~\ref{apxfig:sensor_drift_geometry} shows the assumed pixel geometry and electrostatic elements in the model, but here only depicts a 2$\times$2 pixel region close to the channel. Collected conversions are represented by the four ``bubble'' like structures that hover over positions between pairs of extruded arrows in two dimensional symmetric arrangement about the potential wells. The potential wells (``bubbles'') do not lie in the plane of the front side clock structure because these devices feature a {\it buried channel}. The integrating and barrier clocks are strip-like equipotentials that extend for long distances along the serial address (i) axis and provide boundary conditions that justify utilizing the {method of images} for a small channel depth $z_{chan}$ relative to other relevant dimensions (distance to positions within the drift region, pixel dimension, and the combined width of adjacent integrating clocks). The accumulated conversions constrained to the potential well will appear, in the far field approximation, to have an equal and opposite image charge distribution on the opposite side of the clock plane, acting together as the perturbative, aggressor dipole field denoted $\vec{p}_{ij}$.

Similar arguments are used to describe the far-field influence of the cannel stop ion implants, which, under depleted operation, act as another dipole moment in the $z$ direction with translational invariance along the parallel transfer ($j$) axis. These are denoted $\vec{\xi}_{cs}$ in Figure~\ref{apxfig:sensor_drift_geometry}. Finally, adjacent integrating and barrier clocks act as dipole moments confined to the $i$-$j$ plane with translational invariance along the serial transfer ($i$) axis, shown as $\vec{\xi}_{ck}$.

The predominant component of the drift field is the {backdrop field}, denoted $\vec{E}_{BD}(z)$, is a one dimensional solution of Poisson's equation in the depleted silicon. The influence of the periodic and non-periodic contributors described above can be added in superposition because they explicitly satisfy Gauss' law everywhere except for in those small volumes that contain finite bound charge densities $\rho_b$ not described in a the one-dimensional impurity concentration profile $N(z)$, and surface charge densities $\sigma_f$ arising on semiconductor-conductor interfaces with nonzero normal component of the local electric field. The constant of integration for this backdrop field $\vec{E}_{BD}(z)$ is chosen such that a zero backside bias $BSS$ implies a zero electric field strength directly inside the backside surface of the sensor. In Figure~\ref{apxfig:sensor_drift_geometry} then, the backside window is located a large distance directly above the configuration of electrostatic moments shown, i.e. toward where the nine vertical arrows point. Cold carrier pixel boundaries undergo shifts in response to changes in the positions and magnitudes of the and charge configuration moments $\vec{p}_{ij}$, $\vec{\xi}_{cs,i}$ and $\vec{\xi}_{ck,j}$.

The equations used for vector integration along the drift field lines are:
\begin{eqnarray}
\vec{E}^{tot}(\vec{x}) &=& \vec{E}^{BD}(z) + \delta \vec{E}(\vec{x})\\
\vec{E}^{BD}(z) &=& \left({1 \over \epsilon_0 \epsilon_{Si}} \int^{t_{Si}}_{z} dz N_a (z) - V_{BSS}/t_{Si}\right) \hat{z}\label{eq:ebd}\\
%\label{eqn:drift} \delta \vec{x}_\perp(\vec{x}_0) \cdot \hat{e}_{1,2} &=& \int^{\vec{x}_0}_{\vec{x}\cdot\hat{k}=z_{ch}} d\vec{l} \cdot \hat{e}_{1,2}\\
d \vec{l} &=& {\vec{E}(\vec{x}) \over \left| \vec{E}(\vec{x}) \right|}\, ds \\
\vec{x}_{i+1} &=& \vec{x}_i + d\vec{l}\\
t_{coll}(\vec{x_0}) &=& \int^{\vec{x}_0}_{\vec{x}\cdot\hat{k}=z_{ch}} {dl \over \mu_e\left(E(z),T\right)\, \left| E(z) \right| }.
\end{eqnarray}
where the {\it cold electron} collection time $t_{coll}(\vec{x_0})$ is used to estimate the thermal diffusion at the end of the axial drift, using the mobility in the small field limit, $\sigma^2 \approx 2k_BT/q_e\times\mu_e(E=0,T)\times t_{coll}(\vec{x_0})$. Mobility parameterizations of Jacoboni~\cite{Jacoboni:1977} were used, although we believe there is mounting evidence in X-ray illumination data for these sensors to suggest that the velocity saturation effect for carriers is not as strong as that model provides, for the operating conditions in question.

Details of the electrostatic influence by periodic barriers, denoted $\vec{\xi}_{cs}$ and $\vec{\xi}_{ck}$ in Figure~\ref{apxfig:sensor_drift_geometry}, are contained in the term $\delta\vec{E}(\vec{x})$, and are subdominant for positions far from the channel, $\vec{x}\cdot\hat{z} \gg z_{chan}$, but compete with and can ultimately dominate influence of the backdrop field $\vec{E}^{BD}$ near the channel.\footnote{Any charge configurations near the front side potential wells\cite{Lage_lastkpc_2015} that would motivate carrying higher order terms in a multipole expansion, are not entertained here. We imagine that such terms would include any finite spatial extent in depth and width of the channel stop implant, and any spatial extent in 3 dimensions of the accumulated signal carriers collected in the potential well. Such higher order terms necessarily would have a shorter range of influence ($|\delta{E}|\sim r^{-s},s\ge4$). At a level where they might be important in the drift calculations, these terms will also influence the shapes of charge clouds residing in adjacent wells, complicating the drift calculation. We plan to neglect such terms until there is sufficient evidence in the data to suggest their importance.}  The image charge modeling strategy used, and also the (infinite) periodic arrangement of the channel stop and clock barrier potentials are explicitly given in Rasmussen~\cite{Rasmussen_paccd_2015} [\S\S 3.2-3.3] and are not reproduced here.

\section{Shoelace formulae utilized}\label{apxsec:shoelace}
After pixel boundaries are sampled via the drift calculation, they are compiled into lists that comprise polygonal representations of the pixels. The following formulae were used to compute geometric parameters for each pixel. While direct mapping (e.g., $\vec{x} \in \mathrm{pixel}[i,j]$ {\it vs.} $\vec{x} \ni \mathrm{pixel[i,j]}$) is utilized for certain simulation applications via efficient {\it point-in-polygon} routines, interpretation of recorded images may be aided with use of ancillary pixel information according to the recorded signal distribution in the pixels. Polygon representations of pixels influenced by the recorded signal distributions are straightforward to perform if the signal scale specific perturbation patterns are known. The following shoelace formulae were given in Rasmussen~\cite{Rasmussen_paccd_2015}[\S3.5] that connect pixel area $\mathrm{A}_{ij}$, pixel astrometric shifts $\mathrm{I\textsc{x}}_{ij}$ \& $\mathrm{I\textsc{y}}_{ij}$, and second moments $\mathrm{I\textsc{xx}}_{ij}$, $\mathrm{I\textsc{yy}}_{ij}$ \& $\mathrm{I\textsc{xy}}_{ij}$, given an ordered set of $N$ vertices $(x,y)_{k}$ where $(x,y)_N \equiv (x,y)_0$. For the pixel boundary calculations represented in this work, we typically worked with either 15 or 25 points per side ($60\leq N\leq 100$):

\begin{eqnarray}
\mathrm{A}_{ij}&\equiv& +{1 \over 2} \sum_{k=0}^{N-1} (x_{k+1}y_{k}-x_{k}y_{k+1});\label{eq:aij}\\
\mathrm{I\textsc{xx}}_{ij}\mathrm A_{ij}&\equiv& -{1 \over 12} \sum_{k=0}^{N-1} (y_{k+1}-y_{k})(x_{k}^2 + x_{k+1}^2) (x_{k}+x_{k+1});\\
\mathrm{I\textsc{yy}}_{ij}\mathrm A_{ij}&\equiv& +{1 \over 12} \sum_{k=0}^{N-1} (x_{k+1}-x_{k})(y_{k}^2 + y_{k+1}^2) (y_{k}+y_{k+1});\\
\mathrm{I\textsc{xy}}_{ij}\mathrm A_{ij}&\equiv& +{1 \over 6} \sum_{k=0}^{N-1} (x_{k+1}-x_{k}) x_{k} (y_{k}^2 + y_{k+1}^2 + y_{k}y_{k+1}) \nonumber\\
&&+ {1 \over 24} \sum_{k=0}^{N-1} (x_{k+1}-x_{k})^2 (y_k^2+3y_{k+1}^2+2y_{k}y_{k+1});\\
\mathrm{I\textsc{x}}_{ij}\mathrm A_{ij}&\equiv& -{1 \over 6} \sum_{k=0}^{N-1} (y_{k+1}-y_{k})(x_{k}^2+x_{k+1}^2+x_k x_{k+1});\\
\mathrm{I\textsc{y}}_{ij}\mathrm A_{ij}&\equiv& +{1 \over 6} \sum_{k=0}^{N-1} (x_{k+1}-x_{k})(y_{k}^2+y_{k+1}^2+y_k y_{k+1}).
\end{eqnarray}

The sign of Eq.~\ref{eq:aij} corresponds to a specific choice of chirality for the polygonal vertex list. The quantities above are used to evaluate distortions to pixel area ($\delta \ln \mathrm A_{ij}$), pixel astrometric shift vectors ({\it e.g.}, $\vec{p}_{ij}\cdot\hat{x}=[\mathrm{I}\textsc{x}_{ij}\mathrm{A}_{ij}]/\mathrm{A}_{ij}$) and pixel ellipticities ($\delta \epsilon_{1,ij}=[\mathrm{I}\textsc{xx}_{ij}\mathrm{A}_{ij} - \mathrm{I}\textsc{yy}_{ij}\mathrm{A}_{ij}]/[\mathrm{I}\textsc{xx}_{ij}\mathrm{A}_{ij} + \mathrm{I}\textsc{yy}_{ij}\mathrm{A}_{ij}]$; $\delta \epsilon_{2,ij}=2\,\mathrm{I}\textsc{xy}_{ij}\mathrm{A}_{ij}/[\mathrm{I}\textsc{xx}_{ij}\mathrm{A}_{ij} + \mathrm{I}\textsc{yy}_{ij}\mathrm{A}_{ij}]$). It may be possible for existing pixel data pipelines to be retrofitted to take advantage of such bookkeeping information when estimating object parameters, particularly for PSF estimation purposes. 

\section{A linear perturbation template to represent dynamic pixel response}\label{apxsec:template}

The proportional pixel boundary shifts laid out by Antilogus {\it et al.}\cite{Antilogus_paccd_2014}\S4.2 (and subsequently Refs.~\citenum{Guyonnet_2015,Gruen_jinst_2015}), uses constructions that linearly accumulate the influence of aggressors in the pixel's vicinity. The coupling coefficients are determined using a matrix inversion of constraint equations (containing measured covariances) that utilize reflection symmetries, and a sum rule ({\it e.g.}, Ref.~\citenum{Guyonnet_2015}~\S6.1). Our detailed electrostatic drift calculation may also be applied, and we can do so while explicitly guaranteeing the continuity equation and one-to-one mapping between a two dimensional continuous position field and pixel address. In other words, the Greens function doesn't suffer problems intrinsic to a general arrangement of rectangular pixels that naturally over- and under-claim pixel ``ownership'' of the continuous position field. 

\begin{figure} [ht]
\begin{center}
\includegraphics[width=0.6\linewidth]{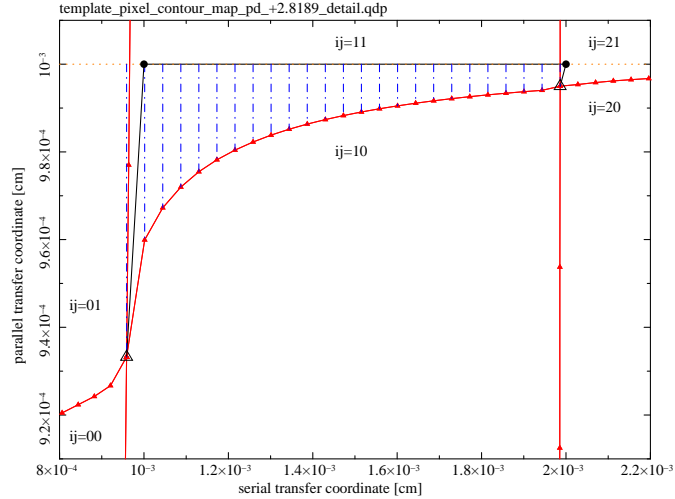}
\end{center}
\caption{
\label{apxfig:template_detail} 
An illustration to show the element level information content of the pixel distortion template. This is for the lower boundary for the pixel with lag $ij=11$ (and contains the same information as the upper boundary for the pixel with lag $ij=10$). 
The (red) filled triangles connected by solid lines form the pixel boundary locus $\vec{x}_{l}^{\,t(i,j)}$ and divide the pixel areas labeled by their corresponding lag ($ij=00,10,20,01,11,21$). The (black) solid line connecting filled circles shows the position of the undistorted pixel boundary (between $ij=10,11$). Open (blue) triangles show positions of the distorted pixel corners, with (black) solid lines ($\delta\vec{c}_{k,k+1}^{\,t(1,1)}$) connecting them to the undistorted corners (filled circles). With distorted corner coordinates projected onto the $\hat{e}_0$ (serial, horizontal) coordinate and the separation divided equally ($n$ samples per side), the $\hat{e}_1$ (parallel, vertical) deflections $\delta d_{l}^{\,t(1,1)}$ of the border are recorded and stored as a template for downstream use. With $n=25$ as shown here, the template can be stored with modest memory requirements, 32 numbers per border per lag, and when compiled for an arbitrary recorded charge distribution, 96 $(x,y)$ pairs per pixel that form a closed polygon. Application of the shoelace formulae (Appendix~\ref{apxsec:shoelace}) provides efficient distillation of this information to 6 leading geometric terms per pixel.
This template was generated for an aggressor amplitude $\bar{p}=2.819\,p_0$: 492 times the effective exposure averaged, aggressor level fit described in Table~\ref{tab:bestfit_pars}. This corresponds to a readout time aggressor signal of $119\mathrm{ke^-}$, comparable to the full well depth for these sensors. 
}
\end{figure}

In the same spirit, we apply the Greens function according to the supposition that all deflections of pixel boundaries are perturbations that scale linearly with aggressor amplitude. We refer to application of the linear perturbation equations collectively as the {\it BF template}. Figure~\ref{apxfig:template_detail} illustrates the geometry. The equations used are as follows, where $\delta\vec{c}_{k,k+1}^{\,t}$ are computed distortion vectors of two adjacent corners for the {\it template aggressor} $p_t$, nominally separated by a single pixel step along the {\it positive} $m$ transfer direction $\hat{e}_{m}$: $m \in \lbrace 0,1 \rbrace$, $\hat{e}_{m} \in \lbrace\left({1 \atop 0}\right),\left({0 \atop 1 }\right)\rbrace$:
\begin{eqnarray}
\delta\vec{x}_l^{\,t} \cdot \hat{e}_{m} & \equiv & \left(\delta\vec{c}_{k}^{\,t} + \left({l \over n-1}\right) \left(s\,\hat{e}_{m}+\delta\vec{c}_{k+1}^{\,t}-\delta\vec{c}_{k}^{\,t}\right)\right)\cdot \hat{e}_{m}\nonumber\\
\delta\vec{x}_l^{\,t} \cdot \hat{e}_{(m+1)\bmod 2} & \equiv & \delta\vec{c}_{l}^{\,t}\cdot  \hat{e}_{(m+1)\bmod 2}+ (\delta d_l^{\,t} - \delta d_0^{\,t}),\nonumber
\end{eqnarray}
where $\delta\vec{x}_l^{\,t}$, $l \in \lbrace 0 \ldots n-1 \rbrace$ are solutions to the electrostatic drift calculation (for template aggressor $p_t$) that form a locus for the pixel boundary of this border, $\delta d_l$ are the boundary distortions perpendicular to the border axis in the $(m+1)\bmod 2$ direction as shown. As usual, $i$ and $j$ are the lag indices in the $m=0$ and $m=1$ directions respectively, and $s$ is the pixel spacing. The per-lag template quantities $\delta\vec{c}_{k}^{\,t(i,j)}$ and $\delta d_l^{\,t(ij)}$ are then compiled by summing over influences of aggressors accumulated at the channel according to:
\begin{eqnarray}
\delta\vec{c}_{k}^{\,(q,r)} &=& \sum_{ij} \left({p_{q-i,r-j} \over p_t}\right)\delta\vec{c}_k^{\,t(i,j)}\nonumber\\
\delta d_{l}^{\,(q,r)} &=& \sum_{ij} \left({p_{q-i,r-j} \over p_t}\right) \delta d_l^{\,t(i,j)}\nonumber\\
\vec{x}_l^{\,(q,r)} \cdot \hat{e}_{m} & \equiv & \left(\left(m\bmod 2 == 0 \right)?q : r \right)\,s + \left(\delta\vec{c}_{k}^{\,(q,r)} + \left({l \over n-1}\right) \left(s\,\hat{e}_{m}+\delta\vec{c}_{k+1}^{\,(q,r)}-\delta\vec{c}_{k}^{\,(q,r)}\right)\right)\cdot \hat{e}_{m}\nonumber\\
\vec{x}_l^{\,(q,r)} \cdot \hat{e}_{(m+1)\bmod 2} & \equiv & \left(\left((m+1)\bmod 2 == 0\right)? q : r \right)\,s + \delta\vec{c}_{l}^{\,(q,r)}\cdot  \hat{e}_{(m+1)\bmod 2}+ (\delta d_l^{\,(q,r)} - \delta d_0^{\,(q,r)}).\nonumber
\end{eqnarray}
In the above, $\delta\vec{c}_k^{\,(q,r)}$ and $\delta d_{l}^{\,(q,r)}$ are the resulting total perturbations to the pixel corners and boundaries from multiple aggressors, respectively; and the boundary pairs $\vec{x}_l^{\,(q,r)}$ for this border are combined the boundary pairs for each of the three other borders to produce a (closed) polygonal description of pixel $(q,r)$. Point-in-pixel algorithms, as well as the shoelace formulae of Appendix~\ref{apxsec:shoelace} are then readily applied to these distorted pixel descriptions. 

The preceding provides a generalized, 2-dimensional application of drift calculation results as linear perturbations. It is analogous to the simpler, perturbative pixel border shifts described in previous work~\cite{Antilogus_paccd_2014,Guyonnet_2015,Gruen_jinst_2015}.

% \subsection{Why}\label{apxssec:why}
% Acknowledgements
\acknowledgements
This material is based upon work supported in part by the National Science Foundation through Cooperative Support Agreement (CSA) Award No. AST-1227061 under Governing Cooperative Agreement 1258333 managed by the Association of Universities for Research in Astronomy (AURA), and the Department of Energy under Contract No. DEAC02-76SF00515 with the SLAC National Accelerator Laboratory. Additional LSST funding comes from private donations, grants to universities, and in-kind support from LSSTC Institutional Members.

% References
\bibliography{spie-astro-tel-instrum_2016_arasmus} % bibliography data in report.bib
\bibliographystyle{spiebib} % makes bibtex use spiebib.bst

\end{document}